%
%
%
%
%

\input harvmac
\input epsf
 
\parindent=0pt
\Title{SHEP 98-12}{\vbox{\centerline{A manifestly gauge invariant exact 
renormalization group}}}

\centerline{\bf Tim R. Morris}
\vskip .12in plus .02in
\centerline{\it 
Department of Physics, University of Southampton,}
\centerline{\it Highfield, Southampton SO17 1BJ, UK}
\vskip .7in plus .35in

\centerline{\bf Abstract}
\smallskip 
In these lectures\footnote{${}^\dagger$}{Lectures 
given at the Workshop on the
Exact Renormalization Group, Faro, Portugal, Sept. 10-12, 1998, to be
published by World Scientific.}
 we describe the construction of
a gauge invariant renormalization group equation
for pure
 non-Abelian gauge theory.  In the process,
a non-perturbative gauge invariant
continuum Wilsonian effective action is precisely defined.
The formulation makes
sense without gauge fixing and thus manifest gauge invariance
may be preserved at all stages.
In the large $N$ limit (of $SU(N)$
gauge theory) the effective action 
simplifies: it may be expressed
through a path integral for
a single particle whose trajectory describes
a Wilson loop. Regularization is achieved with the help of a 
 set of Pauli-Villars fields whose formulation follows naturally
in this picture. Finally, we show how
 the one loop $\beta$ function was computed, 
for the
first time without any gauge fixing.

\vskip -1.5cm
\Date{\vbox{
{hep-th/9810104}
\vskip2pt{October, 1998.}
}
}

\def\ins#1#2#3{\hskip #1cm \hbox{#3}\hskip #2cm}
\def\etc{{\it etc.}\ }
\def\ie{{\it i.e.}\ }
\def\eg{{\it e.g.}\ }
\def\cf{{\it cf.}\ }
\def\viz{{\it viz.}\ }

\def\vv{{\it vice versa}}
\def\aka{{\it a.k.a.}\ }

\def\nonp{non-perturbative}
\def\phi{\varphi}
\def\D{{\cal D}}
\def\p{{\bf p}}
\def\q{{\bf q}}
\def\r{{\bf r}}
\def\k{{\bf k}}
\def\x{{\bf x}}
\def\y{{\bf y}}
\def\z{{\bf z}}
\def\w{{\bf w}}
\def\A{{\bf A}}
\def\Z{{\cal Z}}
\def\C{{\cal C}}
\def\Bb{{\bar B}}
\def\ct{{\tilde c}}
\def\hS{{\hat S}}
\def\defeq{:=}  

\parindent=15pt

\newsec{Introduction and motivation}

\nref\kogwil{K. Wilson and J. Kogut, Phys. Rep. 12C (1974) 75.}
The first lecture was a general introduction to the
exact renormalization group\foot{and coincided with the first 
lecture in ref.\ref\YKIS{T.R. Morris, 
in {\it Yukawa International Seminar '97}, 
Prog. Theor. Phys. Suppl. 131 (1998) 395, hep-th/9802039.}} \kogwil.
In these next two lectures, I wish to report on the formulation of a 
gauge invariant exact renormalization group.
I will take the opportunity to describe the motivation behind
the various constructions used, leaving to the forthcoming paper
a concise exposition of the resulting 
framework\nref\Zak{T.R. Morris, in {\it New Developments 
in Quantum Field Theory},
NATO ASI series 366, (Plenum Press, 1998), 
hep-th/9709100.}\nref\RG{T.R. Morris, 
in {\it RG96}, Int. J. Mod. Phys. B12 (1998) 1343, hep-th/9610012.}
\ref\ym{T.R. Morris, in preparation.}.
Our main motivation for searching for such a structure
was to provide an  elegant
 framework formulated directly in the continuum,
as a first step for non-perturbative analytic approximation
methods. Quite generally such renormalization group methods prove
powerful, as we can see from the many examples reported in this 
workshop,\foot{See also for example the 
reviews \YKIS--\RG. }
and of course there is a clear need for a better \nonp\ understanding
of  gauge theory.
However, several other issues are naturally resolved in the process
of solving this first step. 

In recent years there has been 
substantial progress in solving supersymmetric gauge 
theories \ref\sw{See for example the reviews:
K. Intriligator and N. Seiberg, Nucl. Phys. Proc. Suppl.
45BC (1996) 1, hep-th/9509066;\ C. Gomez and R. Hernandez, 
hep-th/9510023;\ L. Alvarez-Gaum\'e and S.F. Hassan, Fortsch. Phys. 45 (1997)
159, hep-th/9701069.}. These methods involve 
computing a low energy gauge invariant Wilsonian 
effective action, which however, because of the lack of a suitable
framework, is never precisely defined.
Whilst we concentrate here solely on pure Yang-Mills theory, we
see no essential difficulty in generalising the flow equations
 to include (chiral) fermions 
and scalars and indeed supersymmetry.
 It is clear then that our
framework can\foot{At finite $N$, further modification of 
the  Pauli-Villars regularisation may be needed.}
 underpin these ideas.

Whilst it is an interesting academic issue to establish the existence of
a gauge invariant Wilsonian effective 
action and corresponding flow equation,
its use would ultimately be limited without powerful \nonp\ approximation 
schemes. 
Fortunately, a beautiful approximation scheme
lies waiting to be developed: namely 
the large $N$ limit where the gauge group is \eg $SU(N)$
\ref\thooft{G. 't Hooft, Nucl. Phys. B72 (1974) 461.}\nref\mig{A. Migdal, 
Ann. Phys. 109 (1977) 365;\
Yu.M. Makeenko and A.A. Migdal, Nucl. Phys. B188 
(1981) 269.}\nref\LN{A.A. Migdal, Phys. Rep. 102 (1983) 199.}\nref\polbo{
A.M. Polyakov, {\sl Gauge fields and Strings}
(Harwood, 1987).}--\ref\others{V.A. Kazakov and I.K. Kostov, Nucl. Phys.
B176 (1980) 199; V.A. Kazakov, Nucl. Phys. B179 (1981) 283.}.
 Typically, the starting
point for these methods has been  Dyson-Schwinger equations for 
Wilson loops derived at the bare level.
Progress has been hampered by the lack
of corresponding renormalized equations\polbo.
As we saw in the first lecture however, 
one of the most attractive features of the 
exact renormalization group is the fact that solutions may readily be
expressed directly in renormalised terms\YKIS. Thus
combining these two
approaches removes one obstacle
to solving the large $N$ limit. 

We will see that
the large $N$ limit of the Wilson flow equations however results in an
intriguing picture
which is (somehow) dual to the Dyson-Schwinger approach. For example,
in this picture the gauge
fields appear not integrated over but take the r\^ole of background
field `spectators' while the Wilson loop, which is fixed
(but selectable) in the Dyson-Schwinger approach, is now dynamical
and integrated over. The analogue of Migdal's
observation\mig\ that the large $N$
limit may be expressed in terms of the expectation value of a
single Wilson loop\foot{due to correlators of Wilson loops decoupling
in the planar limit\thooft}
is here reflected in the fact that the 
continuum limit of the Wilsonian effective action,
which at finite $N$ may be written in terms of (infinite) sums of 
integrals of products of
Wilson loops, 
at infinite $N$ reduces to a single integral over configurations
of just one Wilson loop. The flow equations reduce to equations 
determining the path integral measure for this Wilson loop.
Operators, which may be viewed as perturbations of this
action, also take the form of averages over 
Wilson loop configurations. These are nothing but the
continuum counterparts of the
`interpolating' operators used in lattice gauge theory to create 
propagating glueball states and study their wavefunctions.
Recall indeed from the first lecture, that the connected Green
functions are already encoded in the effective action. By appropriate
choice of sources, the
same is true here.
This picture is thus ideally suited to describing both
the 
gauge fields and the low energy (\eg bound state) degrees of freedom.

Let us emphasise that in this way our large $N$ limit 
collapses the quantum field theory to a form
of {\sl single particle quantum
mechanics}. By far the most exciting possibility raised by this
viewpoint, in our
opinion, is that it may finally open the door to the solution of
large $N$ gauge theory. (Quite apart from the obvious theoretical
attractions, the large $N$ limit is expected to be accurate 
in practical situations \eg 10\% accuracy or better is expected
for many quantities in
$SU(3)$ Yang-Mills\ref\acc{See the review by A.V. Manohar, in {\sl PANIC 96},
hep-ph/9607484;\ 
M.J. Teper, hep-lat/9804008.}.)
In these lectures, this picture lies just below the surface.
Nevertheless, it was
central in guiding us to the construction of 
a consistent flow equation.

A particular problem that has to be faced in this direction,
is that a gauge invariant
effective cutoff function, similarly to gauge invariant higher derivative
regularisation, is not sufficient to regulate all ultra-violet 
divergences. One loop divergences slip through
\ref\oneslip{A.A. Slavnov,
Theor. Math. Phys. 13 (1972) 1064;\ B.W. Lee and J. Zinn-Justin,
Phys. Rev. D5 (1972) 3121.}.
In standard perturbation theory, this
problem has been cured by supplementing the higher derivative 
regularisation with a system of Pauli-Villars regulator (PV) fields,
the action being bilinear in these fields so that they provide,
on integrating out, the
missing one loop counterterms\foot{and of course other finite
contributions}\ 
\ref\pv{T.D. Bakeyev and A.A. Slavnov, Mod. Phys. Lett. A11 (1996) 1539;\
M. Asorey and F. Falceto, Phys. Rev. D54 (1996) 5290;
Nucl. Phys. B327 (1989) 427.}\ref\warr{B.J. Warr, Ann. Phys. 183 (1988) 1.}. 
This solution is
unwieldy, but worse, the property of being bilinear in the PV
fields is not preserved by the flow: as the gauge field is 
integrated out higher-point PV interactions are generated. 
Instead, we uncover a system of
regulating fields that is more natural from the exact
renormalization group point of view, 
particularly so in the Wilson loop picture.
The resulting
balance of fermionic and bosonic degrees of freedom
which is achieved and the simplicity of the ultra-violet
cancellations, seems to
 hint at a Parisi-Sourlas\ref\ps{G. Parisi and N. Sourlas,
Phys. Rev. Lett. 43 (1979) 744.} 
supersymmetry although, if realised
in spacetime,  of a most unusual kind. 

At the moment
our PV construction is formulated only in the large $N$ limit.
Further adjustments may be needed at finite $N$ (perhaps involving
singlet PV fields), but this investigation is left
for future work.

Last but by no means least, 
the fact that the quantity being calculated,
the effective action, is gauge invariant, suggests
that we ought to be able to dispense with gauge fixing and ghosts. The
gauge fixing procedure
should only be necessary either because the correlators being
calculated require gauge fixing for their definition
or because in the usual way of
performing weak coupling perturbation theory one expands around the
bilinear term $\sim (d\wedge A)^2$ which in itself is not
gauge invariant.\foot{In QED this problem appears through
the fermion bilinear.}
We will see that it is indeed possible to explicitly maintain
the gauge invariance 
at all stages,
resulting in  elegant and highly 
constrained relations. 
Some of  the properties that follow are an essential component 
of our PV regularisation scheme.

Actually, honest non-perturbative approaches to non-Abelian gauge theory
that proceed by gauge fixing, must face up to the challenging problem
of Gribov copies\ref\Grib{V. Gribov, Nucl. Phys. B139 (1978) 1;\
I. Singer, Comm. Math. Phys. 60 (1978) 7;\
See \eg C. Becchi, hep-th/9607181;\ 
P. van Baal, hep-th/9711070;\
M. Asorey and F. Falceto, Ann. Phys. 196 (1989) 209;\
K. Fujikawa, Nucl. Phys. B468 (1996) 355.}. 
Here, these problems are entirely avoided.
Indeed we may turn the issue around, and
use the present formulation, since it is already well defined without
gauge fixing, to investigate explicitly 
the (quantum) consequences
of gauge fixing and Gribov copies directly
in the continuum.

In view of the novelty of the present construction,
a basic test of the formalism is desirable. We computed the 
one loop $\beta$ function, directly in the continuum\YKIS,
using entirely general cutoff functions,\foot{up to some basic
criteria on normalisation and ultraviolet decay rates} and
preserving manifest gauge invariance at all stages. The fact that we
obtained the result $\beta_1=-{11\over3}{N\over(4\pi)^2}$,
independently of the choice of
cutoff functions is encouraging confirmation that the expected
universality of the continuum limit
has been successfully incorporated. The fact that it agrees with the
usual perturbative  result
explicitly confirms that the
one loop $\beta$ function is free from Gribov problems, as expected.

\newsec{Gauge invariance {\it versus} integrating out}

As we saw in the first lecture,
a basic notion of the exact renormalisation group 
(RG)\kogwil\ref\wegho{F.J. Wegner and A. Houghton, Phys. Rev. A8 (1973) 401.}
is the division of momenta into
large and small (according to some effective scale $\Lambda$), the higher
momentum modes being those that are integrated out.
Indeed
it is for this reason that the exact RG is sometimes known as the
``momentum-space'' RG. 
In formulating a \nonp\ RG for gauge theory, we
immediately encounter the problem that this division of momenta into high
and low is inherently at odds with the concept of gauge 
invariance \ref\qap{M. D'Attanasio and 
T.R. Morris, Phys. Lett. B378 (1996) 213.}. This is
easy to appreciate if one considers a homogeneous gauge transformation
$\Omega$ acting on some matter fields $\phi(\x)$:
$$
\phi(\x)\mapsto\Omega(\x)\phi(\x)\quad.
$$
Since in momentum space, $\phi(\p)$ is mapped into a convolution with
the gauge transformation, any division into high and low momenta is
seen not to be preserved by gauge transformations. Therefore, we have
just two options: either 

(a) we break
the gauge invariance, and seek to recover it in the limit that
$\Lambda\to0$, or

(b) we generalise the exact RG.

\noindent
Up until now, it is approach (a) that has been followed almost 
exclusively\foot{
Note that ref.\warr\ constructs a gauge invariant regulator
but the (non-trivial) step to a gauge invariant flow equation is
in fact missing.}
\qap\ref\typeai{
C. Becchi, in {\sl Elementary Particles, Field Theory and Statistical
Mechanics}, (Parma, 1993), hep-th/9607188;\
M. Bonini {\it et al}, Nucl. Phys. B409 (1993) 441, B418 (1994) 81, B421 (1994)
81, B437 (1995) 163;\
U. Ellwanger, Phys. Lett. 335B (1994) 364;\
U. Ellwanger {\it et al}, Z. Phys. C69 (1996) 687;\
M. Reuter and C. Wetterich, Nucl. Phys. B417 (1994) 181, B427 (1994) 291;\
K-I. Aoki {\it et al}, Prog. Theor. Phys. 97 (1997) 479;\
M. Pernici {\it et al}, Nucl. Phys. B520 (1998) 469;\
D.F. Litim and J.M. Pawlowski, Phys. Lett. B435 (1998) 181;\ 
M. Simionato, hep-th/9809004 (see especially its review).}.
Can we follow approach (b)?
As we intimated in ref.\Zak, despite the fact  
that all present forms of the continuous
RG are equivalent to each other (\cf the first lecture), 
there are actually infinitely many other
inequivalent exact RGs (inequivalent 
in the sense that they cannot be mapped into
each other by a quasi-local change of variables). 

A fundamental requirement 
of the {\sl exact} RG that must be preserved, is that the
partition function can be {\sl exactly}
rewritten in terms of some low energy
effective action $S$ and effective cutoff $\Lambda$:
\eqna\zorig
$$\eqalignno{ \Z &= \int\!\!\D\phi\ \e{-S_{bare}} &\zorig a\cr
               &= \int\!\!\D\phi\ \e{-S}\quad, &\zorig b\cr}$$
where in the second line it is understood that modes have been integrated
out but the low energy physics is preserved by replacing the bare action
$S_{bare}$ by an effective action $S$. (We suppress the $\Lambda$ dependence
in $S$ here.)

We can achieve this 
if we ensure that an infinitessimal renormalisation group step ($\Lambda
\mapsto \Lambda-\delta\Lambda$) results in a change in the integrand which
is a total derivative, in other words that 
\eqn\psiflow{{\partial\over\partial\Lambda}\, \e{-S}
={\delta\over\delta \phi}\cdot\left( \Psi\, \e{-S} \right)\quad,}
for some functional $\Psi(\x)$. (We suppress the spatial arguments. The dot 
stands for the spacetime integral -- or `trace', together with a trace
over representation indices.)
The present forms of the exact RG
can be written in this way,
for example Polchinski's \ref\Pol{J. Polchinski, Nucl. Phys. B231 (1984) 269.}:
\eqn\pol{\Lambda{\partial\over\partial\Lambda}S
=-{1\over\Lambda^2}\,{\rm tr}\, c'
\left\{ {\delta S\over\delta\phi}
{\delta S\over\delta\phi}-{\delta^2S\over\delta\phi\delta\phi}
+2\left(c^{-1}\partial^2\phi\right)
{\delta S\over\delta\phi}\right\}}
(where $\phi$ is a scalar field, tr is the trace,
$c(p^2/\Lambda^2)$ is a smooth ultra-violet cutoff profile,
and prime denotes differentiation with respect to its argument),
corresponds to
\eqn\polPsi{\Psi=-{c'\over\Lambda^3}\left(
{\delta S\over\delta\phi} 
+2{c^{-1}}\partial^2\phi\right)}
(up to a vacuum energy term discarded in ref.\Pol). 

Incidentally,
this establishes the fact that all present forms of the exact RG
correspond to reparametrizations of the partition 
function. Indeed, we have by \zorig b\ and \psiflow\ that
\eqn\ch{\delta\Z={\partial\Z\over\partial\Lambda}\delta\Lambda\
=\delta\Lambda\,\int\!\!\D\phi\ \left({\delta\over\delta\phi}\cdot\Psi-
\Psi\cdot{\delta S\over\delta\phi}\right)\,\e{-S}}
is generated by the change of variables
\eqn\repar{\phi(\x)\mapsto \phi(\x)+\delta\Lambda\,\Psi(\x)} 
(the first term in \ch\
arising from the Jacobian, and the second from $S$).

This rather counterintuitive result, that integrating out degrees of
freedom is just equivalent to a change of variables, is allowed
in the continuum because we have an infinite number of degrees of freedom
per unit volume.
After integrating out, the reduced
degrees of freedom can still be put, in a quasilocal way, 
in {\sl one to one} correspondence with the original degrees of freedom.

There are a number of general uses for this observation.
A distinguishing feature of 
\pol\ is its relative simplicity,
however, in certain cases \eg in considering wavefunction renormalisation,
or scaling out of the (marginal) coupling,
several equally appropriate, but inequivalent, choices exist. 
We will not discuss these general cases further but
address all further issues 
to our goal: a gauge invariant flow equation. 

Leaving aside the issue of the ultraviolet regularisation (which will
be solved later),
it easy to write down gauge invariant flow equations which preserve
the partition function \zorig{}:
from \repar\ [or directly from \psiflow] we need only choose a $\Psi$ 
that transforms homogeneously. Clearly however,
although all the exact RGs correspond to reparametrizations $\Psi$,
not all reparametrizations correspond to 
RGs! This is an important issue to which we will return.
For the usual forms of 
continuous RG \pol\ we can appeal, as in the first lecture, directly
to derivations which manifestly show that the flow 
corresponds to elimination of higher momentum modes. But 
for the gauge invariant cases we are interested in, 
we know these alternative
derivations are no longer possible.\foot{Inequivalence to Polchinski's
form can be proved, but is surely to be expected, since otherwise
we would have found a local description of Yang-Mills in terms
gauge {\sl invariant} variables.}

We take our second piece of inspiration from 
ref.\ref\ui{T.R. Morris, Phys. Lett. B357 (1995) 225.}. For
pure $U(1)$ gauge theory there is of course
no conflict with integrating out
because the gauge transformation $A_\mu\mapsto A_\mu+\partial_\mu\omega$
contains no convolution. We can thus derive that the reparametrization
\eqn\uipsi{\Psi_\x^\mu=-{c'\over\Lambda^3}\left({\delta S\over\delta A_\mu}
+2c^{-1}\partial_\nu F_{\nu\mu}\right)\quad,}
which implies
\eqn\uierg{\Lambda{\partial\over\partial\Lambda}S=-{1\over\Lambda^2}{\rm tr}
\,c'\left\{{\delta S\over\delta A_\mu}{\delta S\over\delta A_\mu}
-{\delta^2S\over\delta A_\mu\delta A_\mu}+2\left(c^{-1}\partial_\nu
F_{\nu\mu}\right){\delta S\over\delta A_\mu}\right\}}
(and thus the Legendre form actually used in ref.\ui), 
must correspond to integrating out. Although this pure $U(1)$ case is
trivial, the only continuum solution being the Gaussian 
$S={1\over4}\int\!F_{\mu\nu}c^{-1}F_{\mu\nu}$, there is an important
lesson that can be drawn: {\sl we do not have to fix the gauge}. The 
flow equation and its solution are well defined without gauge fixing.

More or less, to generalise this solution to non-Abelian gauge theory, we only
need now to covariantize \uipsi, replacing partial differentials with
covariant differentials. It is helpful to specialize however
to a particular way
of covariantizing these equations (one that was motivated by `pulling apart'
the two $\delta/\delta A_\mu$'s in \uierg\ as much as possible so as to
improve its ultraviolet (UV) convergence properties).

\newsec{A gauge invariant exact RG}

We write the gauge field $A_\mu(\x)=
A_\mu^a(\x)\tau^a$ always contracted into
the generators,
which are taken anti-Hermitian and in the fundamental
representation. These are orthonormalised as 
${\rm tr}(\tau^a\tau^b)=\half\delta^{ab}$.
Motivated by the completeness relation for the $\tau$'s we define
$${\delta\over\delta A_\mu(\x)}\defeq2\tau^a{\delta\over\delta A_\mu^a}\quad.$$
We write the covariant derivative as 
$D_\mu=\partial_\mu-iA_\mu$, and
thus the field strength as \break
$F_{\mu\nu}=i[D_\mu,D_\nu]$.
This is important because exact preservation of the gauge invariance
\eqn\gaugetr{\delta A_\mu=D_\mu\cdot\omega
\defeq[D_\mu,\omega]}
ensures that the gauge field so defined,
cannot renormalise. We will throughout require, as a basic tenet of the
Wilsonian RG, that the effective action $S$ is quasilocal, by which here
we merely mean that there always exists a derivative 
expansion (to all orders).
Quasilocality plus gauge invariance then implies that 
\eqn\defg{S={1\over2g^2}\,{\rm tr}\!\int\!\!d^D\!x\, F_{\mu\nu}^2
+O(\partial^3)\quad,}
where the coefficient of the lowest order -- $O(\partial^2)$ -- term serves
to define the coupling constant
$g$. (We work in Euclidean space of general dimension $D$.)
Note thus that, around the Gaussian fixed point,
since there will be no gauge fixing parameter and no ghosts,
only $g$ will require renormalisation!

We introduce the Wilson line, the path ordered exponential along some curve
$\C_{\x\y}$:
\eqn\defWl{\Phi[\C_{\x\y}]=P\exp-i\int_{\C_{\x\y}}\!\!\!\!\!\!dz^\mu
A_\mu(z)\quad,}
which behaves as a parallel transport factor (for the fundamental
representation). This will be drawn as a thick line with an arrow indicating
the direction of index flow, as in fig.1. Where we have occasion to expand
in the gauge field, as in
\eqn\defWlex{\Phi=1-i\int_0^1\!\!\!\!d\tau\,{\dot z}^\mu\! A_\mu(z)
-\int_0^1\!\!\!\!d\tau_2\!\int_0^{\tau_2}\!\!\!\!\!d\tau_1\,
{\dot z}\!\cdot\! A(\tau_1)\,
{\dot z}\!\cdot\! A(\tau_2)\ +\cdots}
(where we have parametrized the curve by $z^\mu(\tau)$, 
$\tau\in[0,1]$, $\z(0)=\x$,
$\z(1)=\y$), we 
will draw it as a thin line and indicate by blobs the position of the fields,
as also shown in fig.1.
$$
\epsfxsize=\hsize\epsfbox{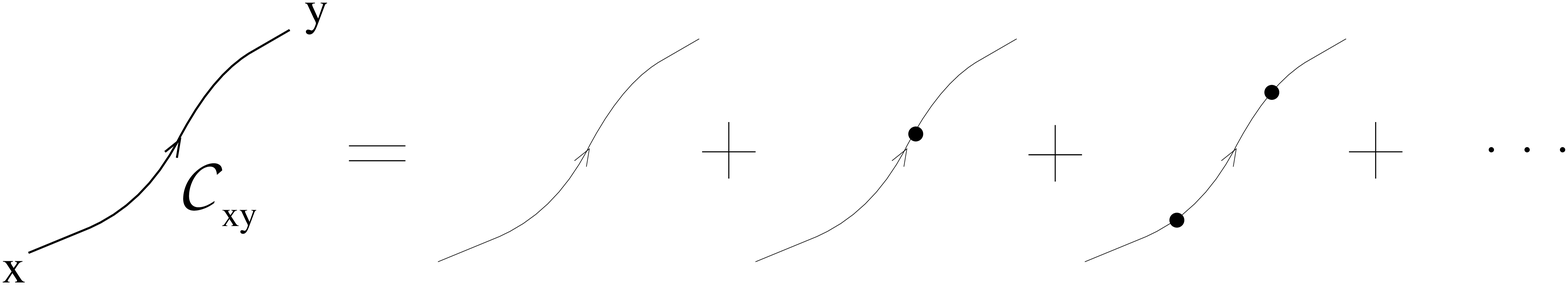}
$$
\centerline{ {\bf Fig.1.} Diagrammatic representation of a Wilson line
and its expansion.}

A useful representation for gauge covariantization,
will be in terms of a path
integral over paths $\C_{\x\y}$ from $\x$ to $\y$.
For two (group singlet) functionals $X$ and $Y$ we write
\eqn\defao{a_0[X,Y]={1\over2\Lambda^2}
\int\!\!d^D\!x\,d^D\!y\int\!\!\D\,\C_{\x\y}\
{\rm tr}\left\{ 
{\delta X\over\delta A_\mu(\x)}\Phi[\C_{\x\y}]
{\delta Y\over\delta A_\mu(\y)}\Phi^{-1}[\C_{\x\y}]\right\}\quad,}
The path integral measure is normalised by
\eqn\aonorm{\int\!\!\D\,\C_{\x\y}\ 1 =c'_{\x\y}\defeq
\int\!\!d^D\!x\,c'(p^2/\Lambda^2)\,{\rm e}^{i\p.(\x-\y)}\quad,}
but is otherwise arbitrary. 
Geometrically, a particularly simple choice
would be to take only one path, the straight line between $\x$ and $\y$,
in which case the measure is just multiplication by $c'_{\x\y}$. 
Analytically, a nicer choice follows 
from the covariantization
\eqn\altdefao{a_0[X,Y]={1\over2\Lambda^2}\int\!\!d^D\!x\ {\rm tr}\left\{
{\delta X\over\delta A_\mu(\x)}c'(-D^2/\Lambda^2)\cdot
{\delta Y\over\delta A_\mu(\x)}\right\}\quad.}
Note the restriction in \defao\ to a pair of coincident lines of 
parallel transport, as represented in fig.2. 
$$
\epsfxsize=0.25\hsize\epsfbox{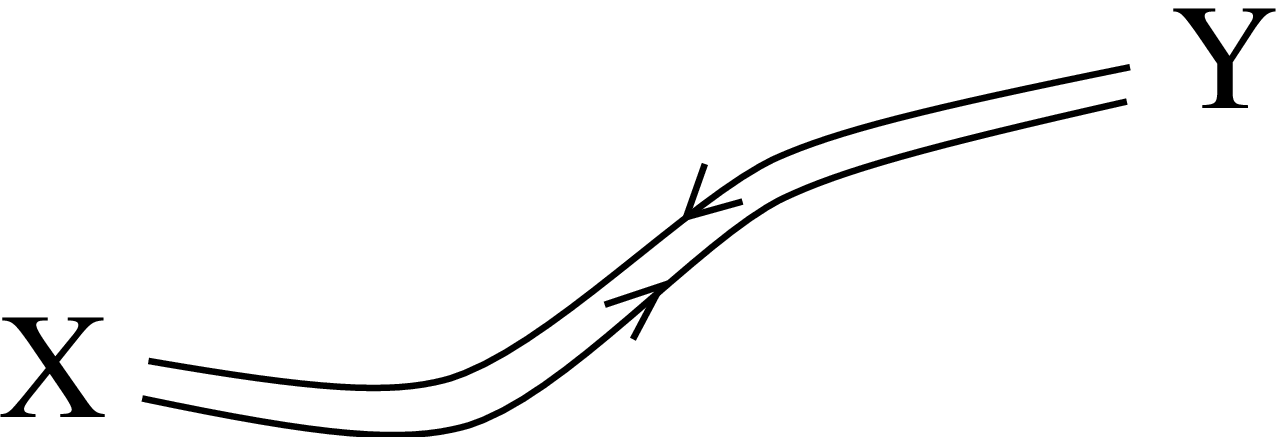}
$$
\centerline{ {\bf Fig.2.} The `wine'.}
(One should understand
that functional derivatives sit in the open ends of the pair.) 
 This choice is for convenience:
when $X$ and $Y$ are  path ordered exponentials, 
we have the $SU(N)$ identity given in fig.3, as follows from completeness
of the $\tau$'s. 
$$
\epsfxsize=.9\hsize\epsfbox{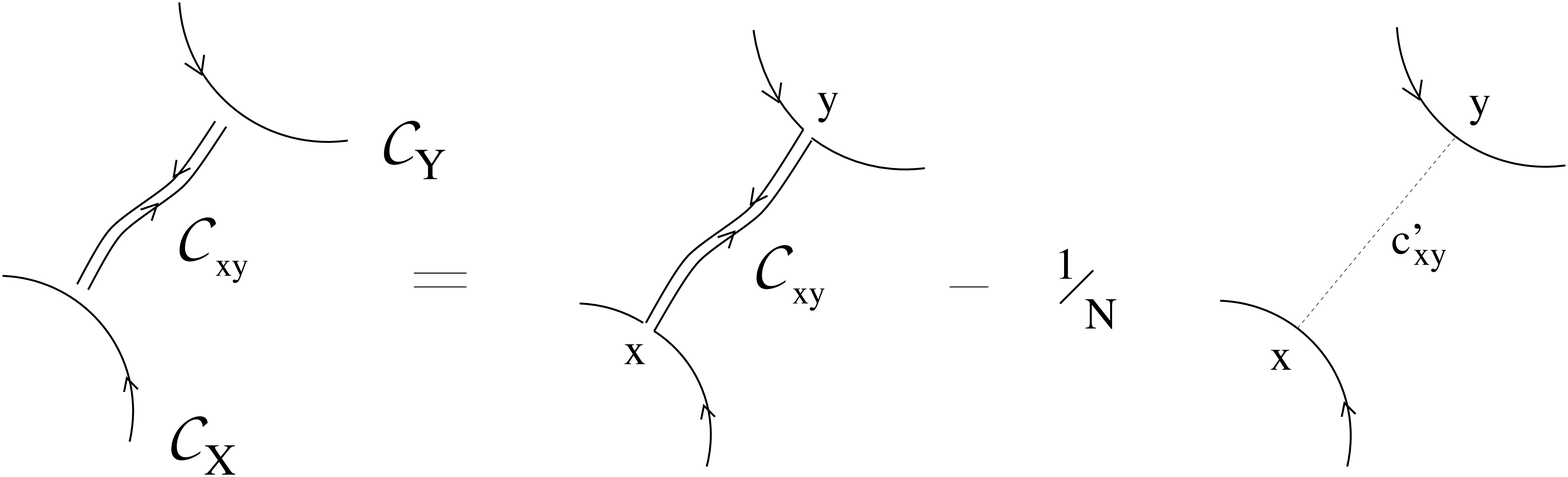}
$$
{\bf Fig.3.} The completeness identity. 
On the r.h.s. $\x$ and $\y$ are integrated round
the curves using $\int_{\C_X}\!\!\!dx^\mu
\int_{\C_Y}\!\!\!dy^\mu$, and the dotted line corresponds simply
to multiplication by $c'_{\x\y}$.
$$
\epsfxsize=0.55\hsize\epsfbox{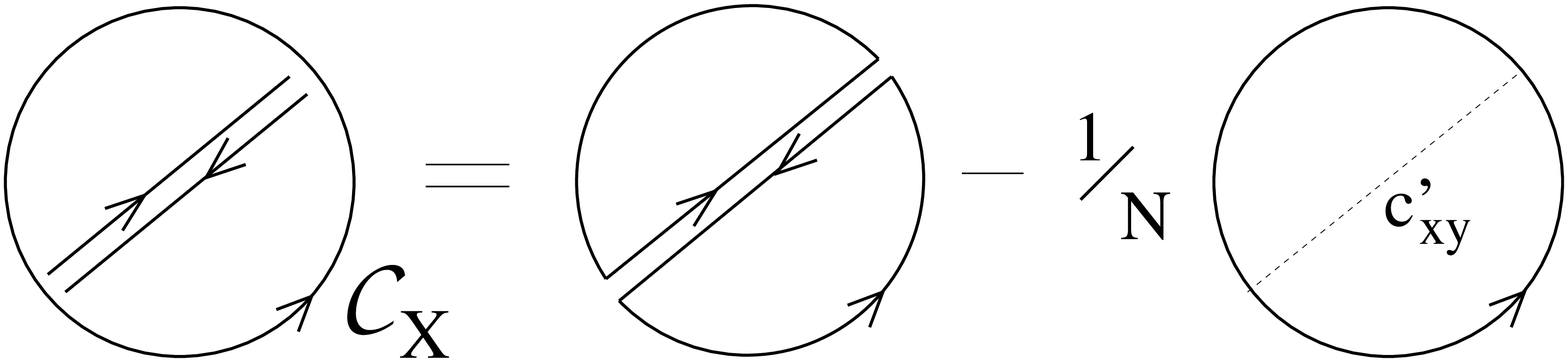}
$$
\centerline{ {\bf Fig.4.} Graphical definition of $a_1[X]$.}
Similarly, we define $a_1[X]$,
where both derivatives act on the same functional as illustrated in
fig.4.  Since the path integral over a
 pair of coincident lines of parallel transport will play
a central r\^ole in the ensuing discussion,  
I will refer to this unwieldy description simply as a `wine'
(from the concatination W[ilson l]ine).

We can now introduce the gauge field sector of the flow equation:
\eqn\erg{\Lambda{\partial\over\partial\Lambda}S=
-a_0[S,g^2S-2\hS]+a_1[g^2S-2\hS]+\cdots\quad,}
where $\hS$ is a gauge invariant `seed' action that we choose, and the dots
stand for PV contributions which will be discussed in the second lecture.
This corresponds to 
$$\Psi^\mu_\x=-{1\over2\Lambda^2}\int\!\!d^D\!y\int\!\!\D\,\C_{\x\y}\
\Phi[\C_{\x\y}]
\left(g^2{\delta S\over\delta A_\mu(\y)}-2{\delta\hS\over\delta A_\mu(\y)}
\right)\Phi^{-1}[\C_{\x\y}]$$ 
and thus exact preservation of the partition function:
$\partial\Z/\partial\Lambda=0$. Actually, in  expanding 
${\delta\over\delta A}(\Psi e^{-S})$, there is in principle
a term where the `wine bites its own tail' as in
fig.5. 
These contributions have unpleasant UV properties. We could discard
these if we regularise the wine so that \eg the path integral excludes paths
$\C_{\x\y}$ which stray inside spheres of radius $\epsilon$ surrounding
the gauge field differentials $\delta/\delta A(\x)$ and  
$\delta/\delta A(\y)$.
But such a regularization temporarily breaks the gauge invariance,
and the limits $\Lambda_0\to\infty$ (continuum limit) and $\epsilon\to0$
need not (and in general do not) commute. Remarkably, it will turn out
that our PV contributions precisely cancel all such terms.
$$
\epsfxsize=0.28\hsize\epsfbox{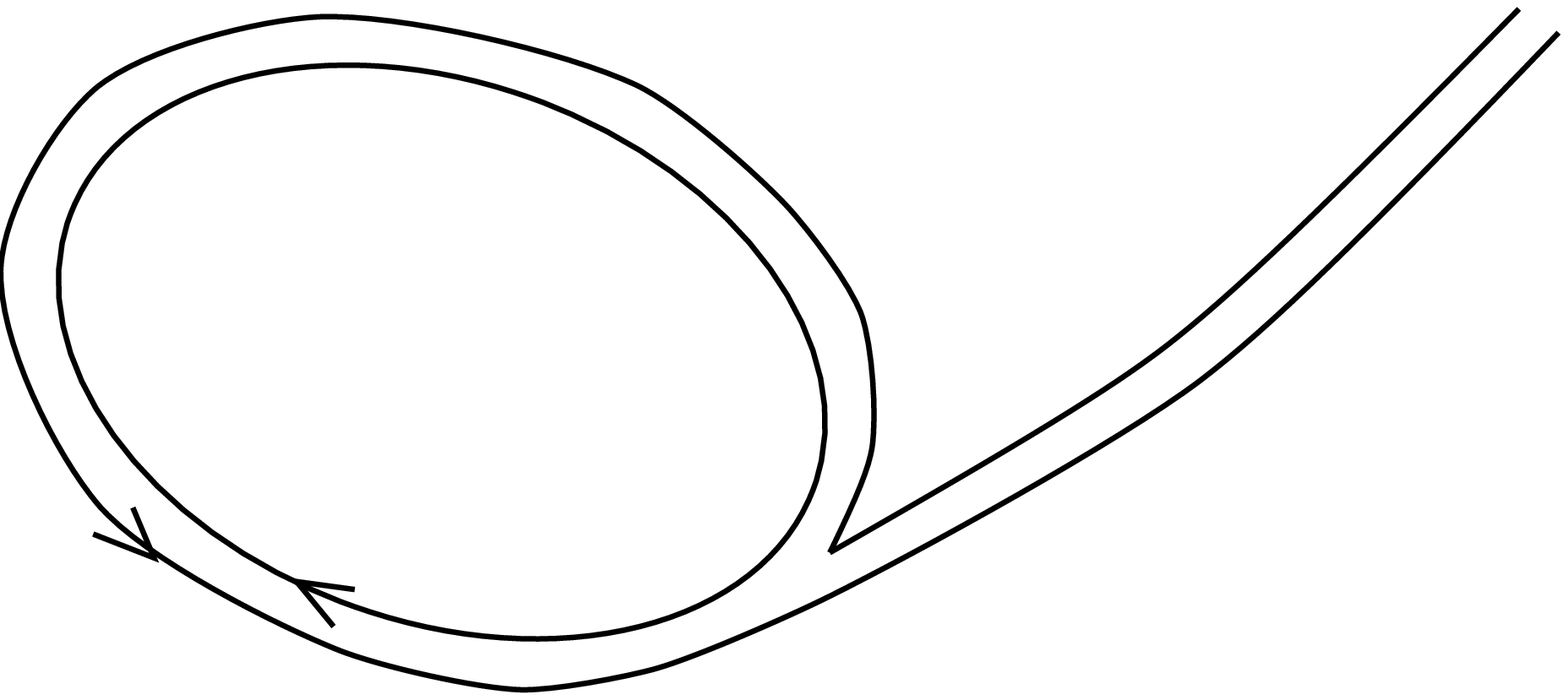}
$$
\centerline{ {\bf Fig.5.} A wine biting its own tail.}

Now note that $S$, and $\hS$, have an expansion in Wilson loops:
\eqn\defWS{S=\int\!\!\D\,\C\,\phi[\C]+{1\over2!}\int\!\!\D[\C_1,\C_2]\,
\phi[\C_1]\,\phi[\C_2]\ +\cdots,}
(where we have introduced, as in fig.4, the Wilson loop $\phi[\C]={\rm tr}\,\Phi[\C_{\x\x}]$,
$\C$ being a closed contour, $\x$ some arbitrary point on it, and
$\C_{\x\x}$ being the marked contour starting and finishing at $\x$).
While this statement might seem surprising, all we are in fact asserting
is that $S$, being a group invariant, has an expansion in traces and 
products of traces:
\eqn\sexp{
\eqalign{S= &{1\over2}\int\!\!d^D\!x\,d^D\!y\, S_{\mu\nu}(\x,\y)\,{\rm tr}\,
A_\mu(\x)A_\nu(\y)\cr
+&{1\over3}\int\!\!d^D\!x\,d^D\!y\,d^D\!z\,
 S_{\mu\nu\lambda}(\x,\y,\z)\,{\rm tr}\,
A_\mu(\x)A_\nu(\y)A_\lambda(\z)\cr
+&{1\over4}\int\!\!d^D\!x\,d^D\!y\,d^D\!z\,d^D\!w\,
 S_{\mu\nu\lambda\sigma}(\x,\y,\z,\w)\,{\rm tr}\,
A_\mu(\x)A_\nu(\y)A_\lambda(\z)A_\sigma(\w)\cr
+&\cdots\cr
+&{1\over2!}\left({1\over2}\right)^2\!\int\!\!d^D\!x\,d^D\!y\,d^D\!z\,d^D\!w\,
 S_{\mu\nu,\lambda\sigma}(\x,\y;\z,\w)\,{\rm tr}\,
A_\mu(\x)A_\nu(\y)\, {\rm tr}\,A_\lambda(\z)A_\sigma(\w)\cr
+&\cdots\quad.}}
We will require of the path integrals whatever sums and measures are
necessary to reproduce these $n$-point vertices.
As far as these lectures are concerned, the path integral over Wilson
loops merely
acts as a mnemonic which summarises the expected properties
of the $n$-point vertices (much as the functional integral should 
strictly be regarded
as doing for $n$-point Green functions). For example, 
we have that the one-Wilson-loop $n$-point
vertices in momentum space are given by
\eqn\npoint{\eqalign{(2\pi)^D &\delta(\p_1+\cdots+\p_n)\,
S_{\mu_1\cdots\mu_n}(\p_1,\cdots,\p_n) =\cr
&(-i)^n\int\!\!\D\,\C\mathop{\oint\!\oint\!\cdots\oint}_{\displaystyle
(1,2,\cdots,n)}\!  dx_1^{\mu_1}dx_2^{\mu_2}\cdots
dx_n^{\mu_n}\ {\rm e}^{-i(\x_1.\p_1+\cdots+\x_n.\p_n)}\quad,} }
where the notation $(1,2,\cdots,n)$ stands for integrating round the loop
while preserving the cyclical order, as illustrated in fig.6. 
$$
\epsfxsize=0.3\hsize\epsfbox{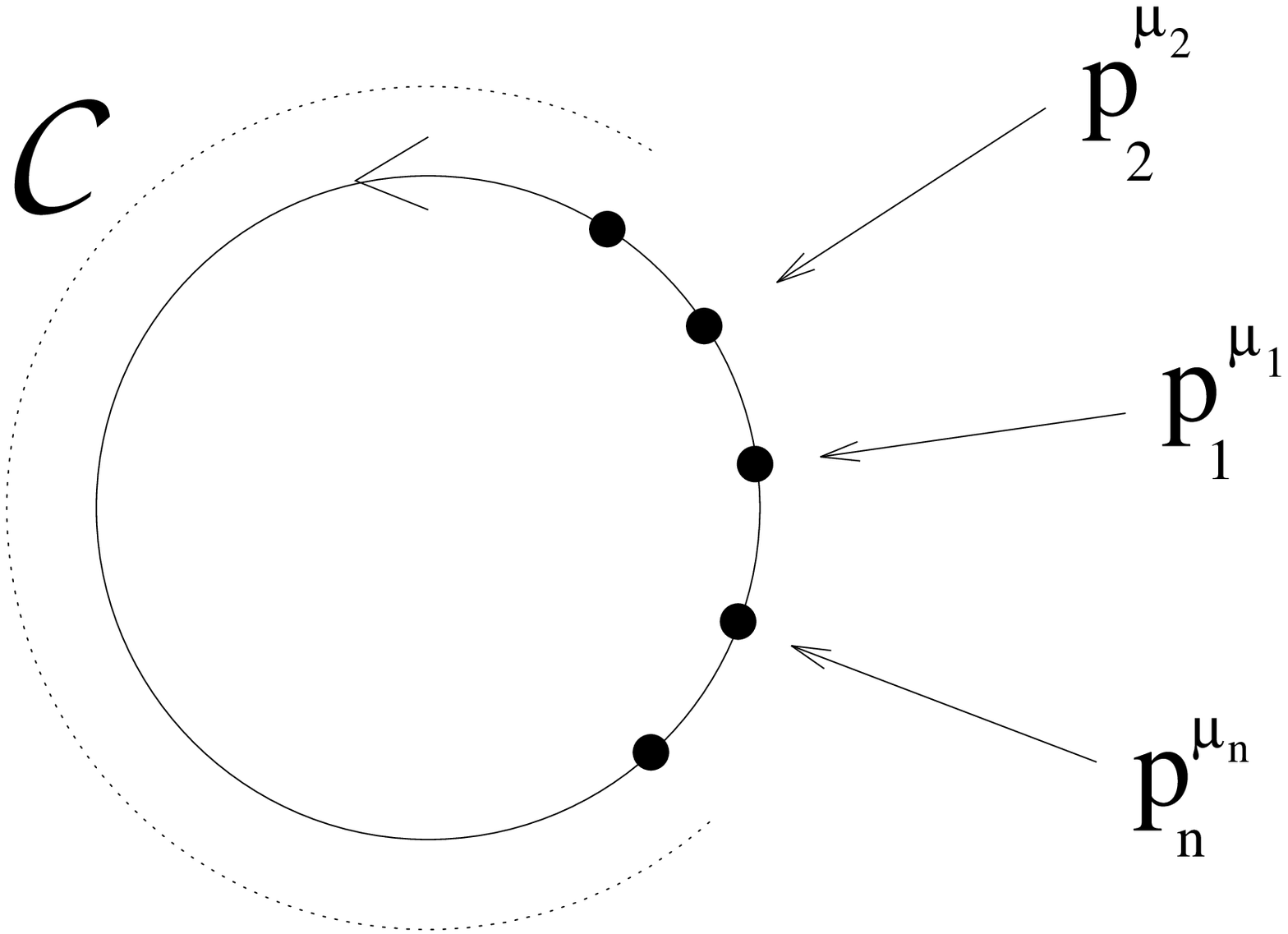}
$$
\centerline{ {\bf Fig.6.} One-Wilson-loop $n$-point vertex.}
From this it
is clear that these $n$-point vertices satisfy cyclicity:
$$S_{\mu_1\cdots\mu_n}(\p_1,\cdots,\p_n)=
S_{\mu_2\cdots\mu_n\mu_1}(\p_2,\cdots,\p_n,\p_1)\quad,$$
charge conjugation invariance ($A_\mu\mapsto-A^T_\mu$ which corresponds
to reversal of the direction of the Wilson loop $\C$):
$$S_{\mu_1\cdots\mu_n}(\p_1,\cdots,\p_n)=
(-)^nS_{\mu_n\cdots\mu_1}(\p_n,\cdots,\p_1)\quad,$$
and gauge invariance:
\eqn\ga{p_1^{\mu_1}S_{\mu_1\cdots\mu_n}(\p_1,\cdots,\p_n)=
S_{\mu_2\cdots\mu_n}(\p_1+\p_2,\p_3,\cdots,\p_n)
-S_{\mu_2\cdots\mu_n}(\p_2,\cdots,\p_{n-1},\p_n+\p_1),}
as follows from $\delta A_\mu=\partial_\mu\omega+[A_\mu,\omega]$, or
simply from direct integration of \npoint. This
corresponds to a `push forward' minus a  `pull back' of the point concerned,
as in fig.7.
$$
\epsfxsize=.63\hsize\epsfbox{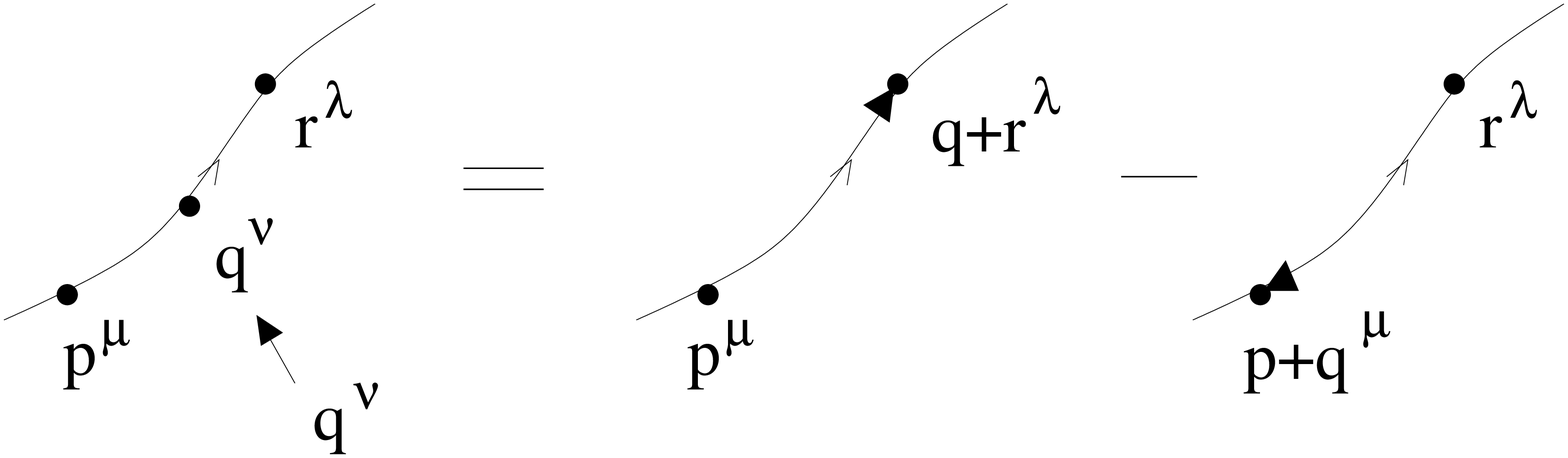}
$$
\centerline{ {\bf Fig.7.} Graphical representation of gauge invariance
identities.}

While at this stage we could dispense with the
 Wilson loop picture -- formulating everything 
directly in terms of vertices, we will see that, as the
lectures progress, the picture becomes more and more
central to understanding the properties of these equations
and indeed will act as a guide to their complete construction.

Now we address the question: what is $\hS$? In our $U(1)$ case we have
\eqn\hsui{
g^2S=\hS={1\over4}\int\!\!d^D\!x\, F_{\mu\nu}\,c^{-1}(-\partial^2/\Lambda^2)
F_{\mu\nu}\quad,}
 \ie the free Gaussian fixed point. In the non-Abelian case
obviously, even at tree level, these relations are no longer possible. At
first sight there appear to be
 many ways to generalise, for example we might 
insist that the first relation
$\hS=g^2S$ remains true at the classical level (the $g\to0$ 
limit).
This fixes $\hS$ through the classical limit of \erg. We could then 
determine the full quantum $S$ through the full flow. Or else we could
fix $\hS$ to be a given covariantization of the r.h.s. of \hsui. We
would lose the relation $g^2S=\hS$ even at the classical level, but
all of $S$ would be determined through \erg. 

It turns out that it has to be the latter. To see this, we note that in fact,
at least at the classical level, we may solve \erg\ for a quasilocal $\hS$
in terms of any quasilocal $S$ of our choosing! Thus without further 
constraints on $\hS$ we can put in whatever physics (or otherwise) that
we wish. The problem is that $\hS$ in these cases has hidden within it
infrared regulated propagators, which at low momenta can always
(by quasilocality) be Taylor expanded in $p^2/\Lambda^2$, 
but at high momenta show up as 
negative powers $\Lambda^2/p^2$ in the asymptotic expansion. 
Therefore we need to require that $\hS$ 
does not hide any infrared regulated propagators, \ie the vertices
do not have any negative powers in their asymptotic expansion.
In fact to be completely
safe we will insist on ultralocality \ie that $\hS$
contains only a {\sl finite} sum of derivative terms. We
achieve this by setting $c^{-1}(x)$ to be a polynomial in $x$ of degree
$r$, and define
\eqn\defhS{\hS={1\over2}{\rm tr}\,\int\!\!d^D\!x\,F_{\mu\nu}(\x)\,
c^{-1}\!\left(-{D^2\over\Lambda^2}\right)\cdot F_{\mu\nu}(\x)\quad.}
More generally we could insert a wine normalised to 
$c^{-1}_{\x\y}\equiv c^{-1}(-\partial^2/\Lambda^2)\delta(\x-\y)$.
The
 extra freedom in the path integral measure amounts to allowing
other higher order interactions (with 
a maximum number of $2r$ derivatives).
Note that such an $\hS$ may thus be represented as a path integral 
$\int\!\D\,{\hat\C}$ over a single Wilson loop, as illustrated in fig.8.
$$
\epsfxsize=.38\hsize\epsfbox{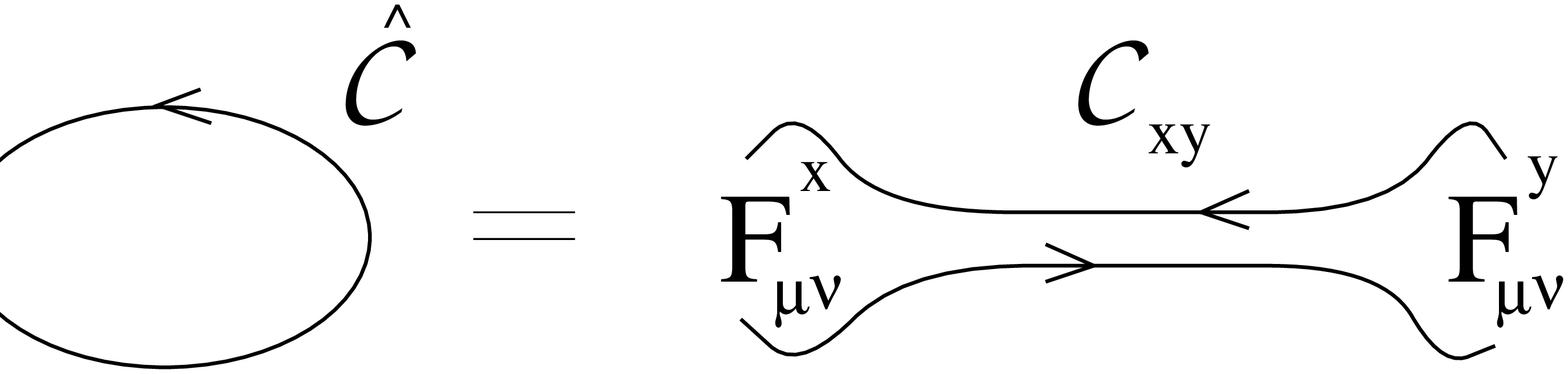}
$$
\centerline{ {\bf Fig.8.} The Wilson loop for $\hS$.}
\vfill\eject

Its coefficient functions follow straightforwardly:
\eqn\hSex{\eqalign{\hS_{\mu\nu}(p) &=2\Delta_{\mu\nu}(p)/c_p\quad,\cr
\hS_{\mu\nu\lambda}(\p,\q,\r) &={2\over c_p}(
p_\lambda\delta_{\mu\nu}-p_\nu\delta_{\lambda\mu})
-2(p.r\delta_{\lambda\mu}-p_\lambda r_\mu)c^{-1}_\nu(\q;\p,\r)
+{\rm cycles}\quad, } }
\etc,
where in the two-point vertex we set $\p_1=-\p_2=\p$, introduce
the shorthand $c_p\equiv c(p^2/\Lambda^2)$
and the transverse combination
$\Delta_{\mu\nu}(p) :=p^2\delta_{\mu\nu}-p_\mu p_\nu$, and
in the three-point vertex we introduce the one-point wine as in fig.9, and
add the two cyclic permutations of $(p_\mu,q_\nu,r_\lambda)$.
For the covariantization implied in \defhS, 
we have\foot{Note, you can see immediately that the gauge transformation
$q^\nu c^{-1}_\nu(\q;\p,\r)$ is as expected by fig.7.
From this, one can check that the gauge transformations of \hSex\ are
as given in \ga.}
\eqn\wone{c^{-1}_\nu(\q;\p,\r)=\,(p-r)^\nu\,{c^{-1}_p-c^{-1}_r
\over q.(p-r)}\quad.}
$$
\epsfxsize=.6\hsize\epsfbox{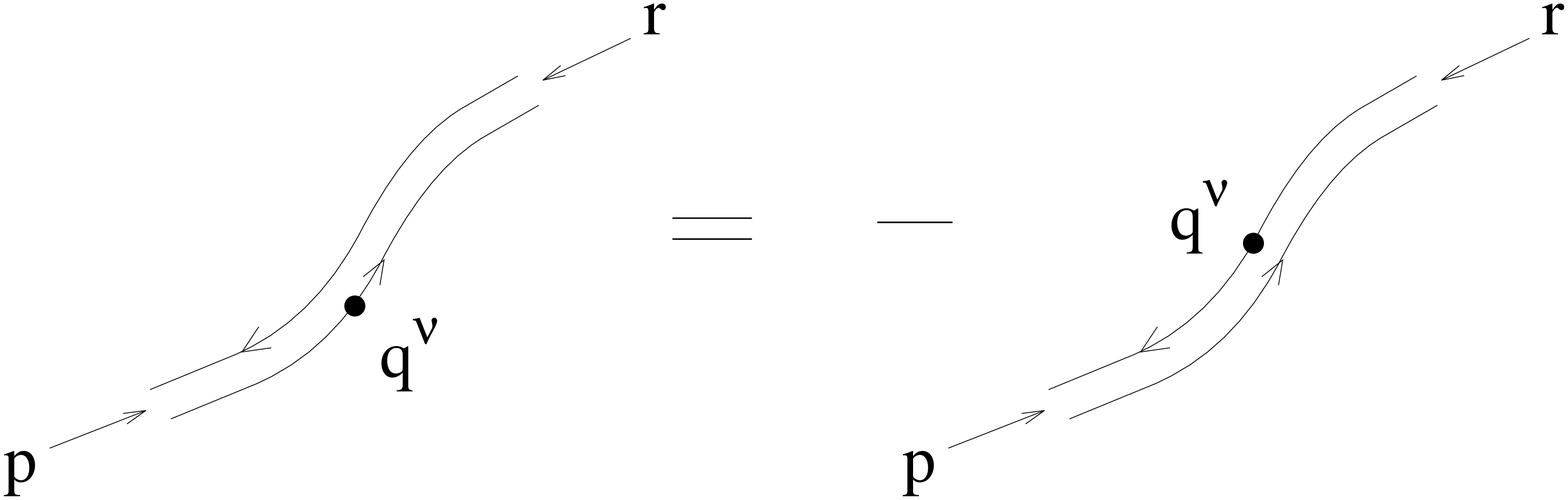}
$$
\centerline{ {\bf Fig.9.} The one-point wine $c^{-1}_\nu(\q;\p,\r)$.}

The full flow equation \erg\ may now be written in Wilson loop
representation, as illustrated in fig.10.
Immediately, we see that
in the large $N$ limit, only a single Wilson loop survives. This
follows through the explicit $1/N$'s in fig.10, and because  
in the large $N$ limit all the gauge fields lie on one half of the
threaded loops, as in fig.11.
Thus in this limit
the continuum solution for $S$ may be expressed as a path integral
over a single particle circulating in a loop:
$$S=P\int\!\!\D x\ {\rm e}^{-s[\x]+\oint\! d\x.\A}\quad,$$
where we have parametrized the closed curve $\C$ by $x^\mu(\tau)$, and
the path integral measure ${\rm e}^{-s}$ is determined by the
flow equation. Gauge theory in the large $N$ limit, is the quantum
mechanics of a single particle!

$$\eqalign{&\Lambda{\partial\over\partial\Lambda}S=\cr
&-g^2\int\!\!\D\,\C_1\D\,\C_2\left\{{\delta S\over\delta\phi[\C_1]}
{\delta S\over\delta\phi[\C_2]}-{\delta^2S\over\delta\phi[\C_1]
\delta\phi[\C_2]}\right\}
\left\{\vcenter{\epsfxsize=0.36\hsize\epsfbox{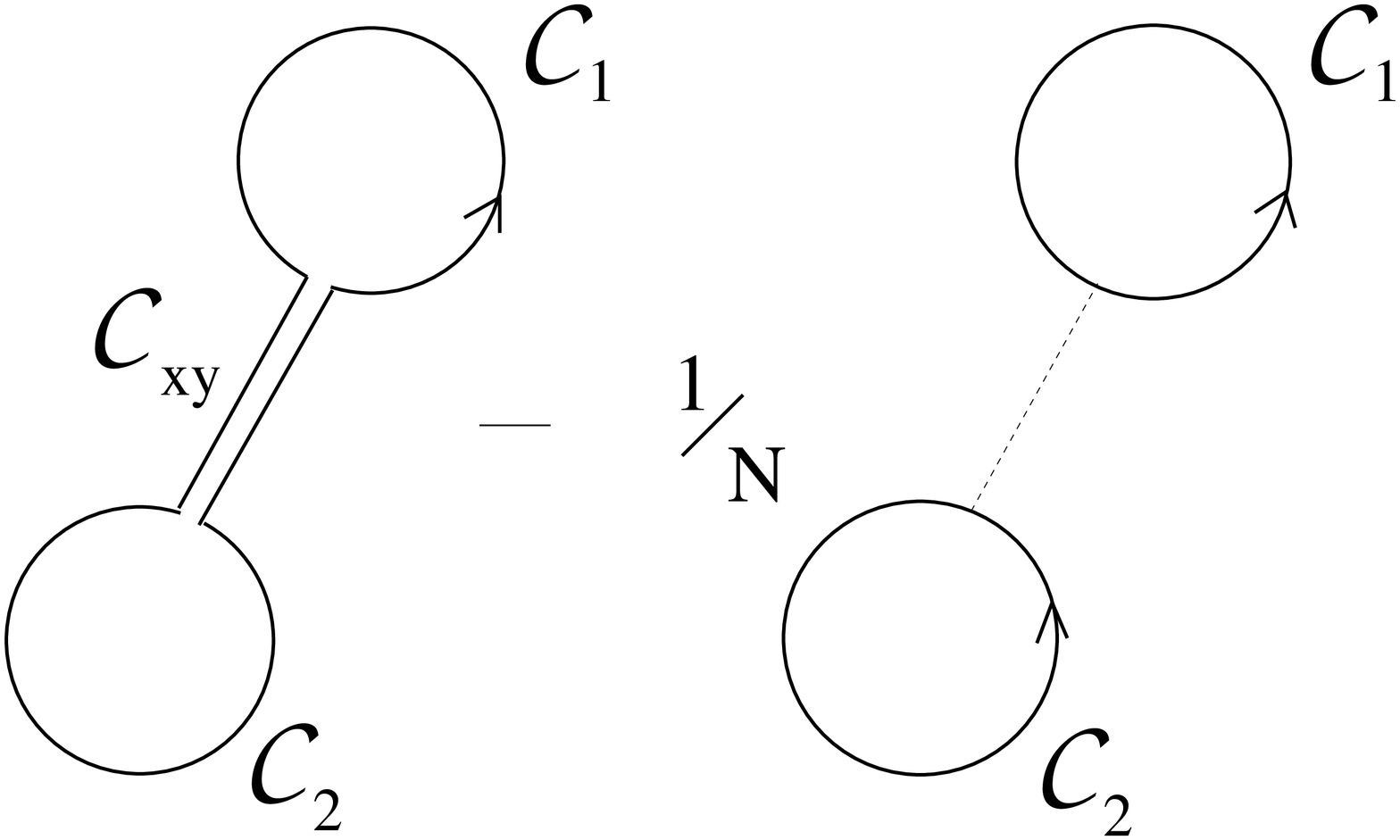}}\right\}\cr
+&2\int\!\!\D\,{\hat\C}\,\D\,\C\,{\delta S\over\delta\phi[\C]}
\left\{\vcenter{\epsfxsize=0.36\hsize\epsfbox{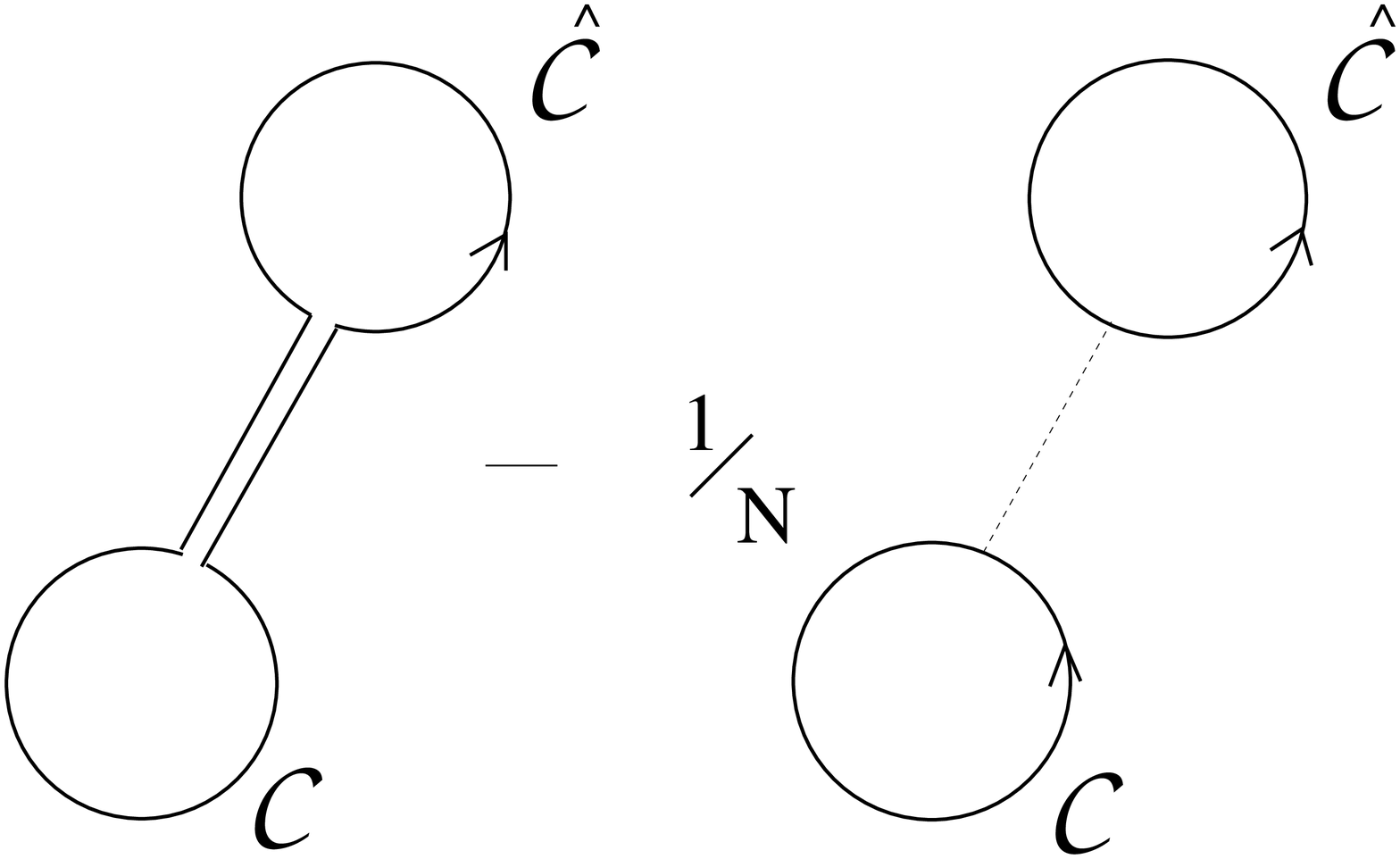}}\right\}\cr
+&g^2\int\!\!\D\,\C\,{\delta S\over\delta\phi[\C]} \left\{
\vcenter{\epsfxsize=0.3\hsize\epsfbox{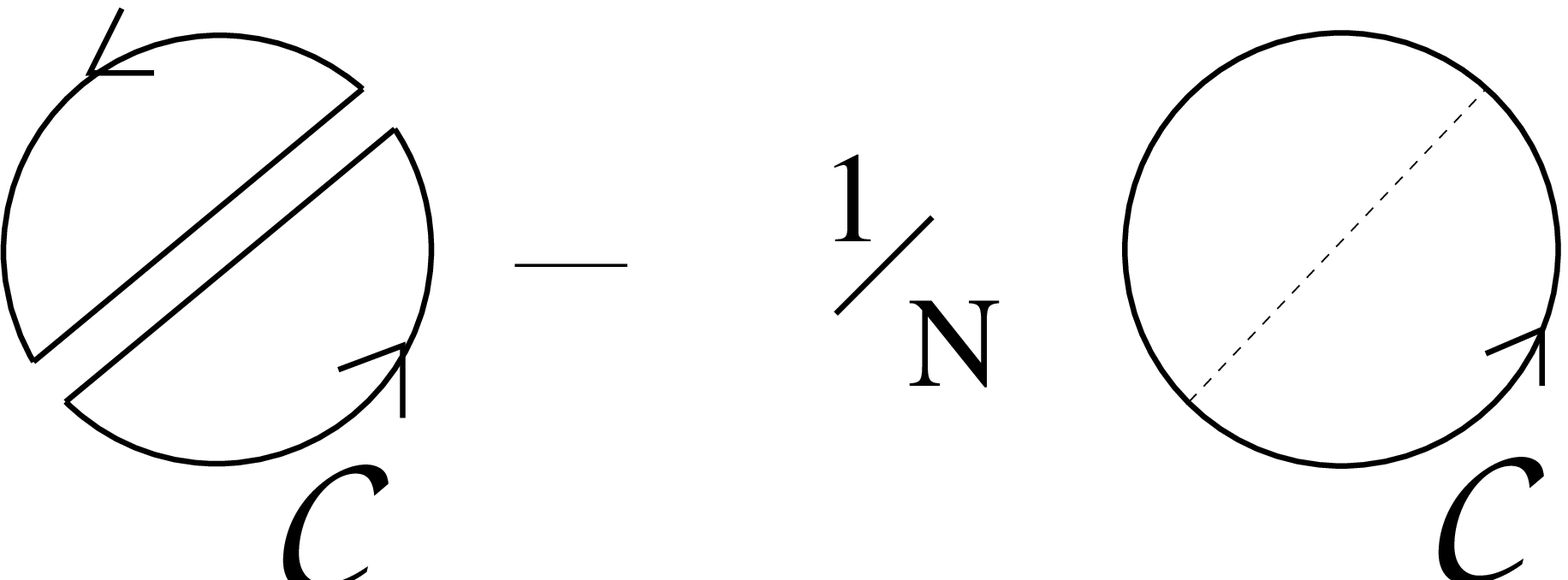}}\right\}
-2\int\!\!\D\,{\hat\C} \left\{
\vcenter{\epsfxsize=0.3\hsize\epsfbox{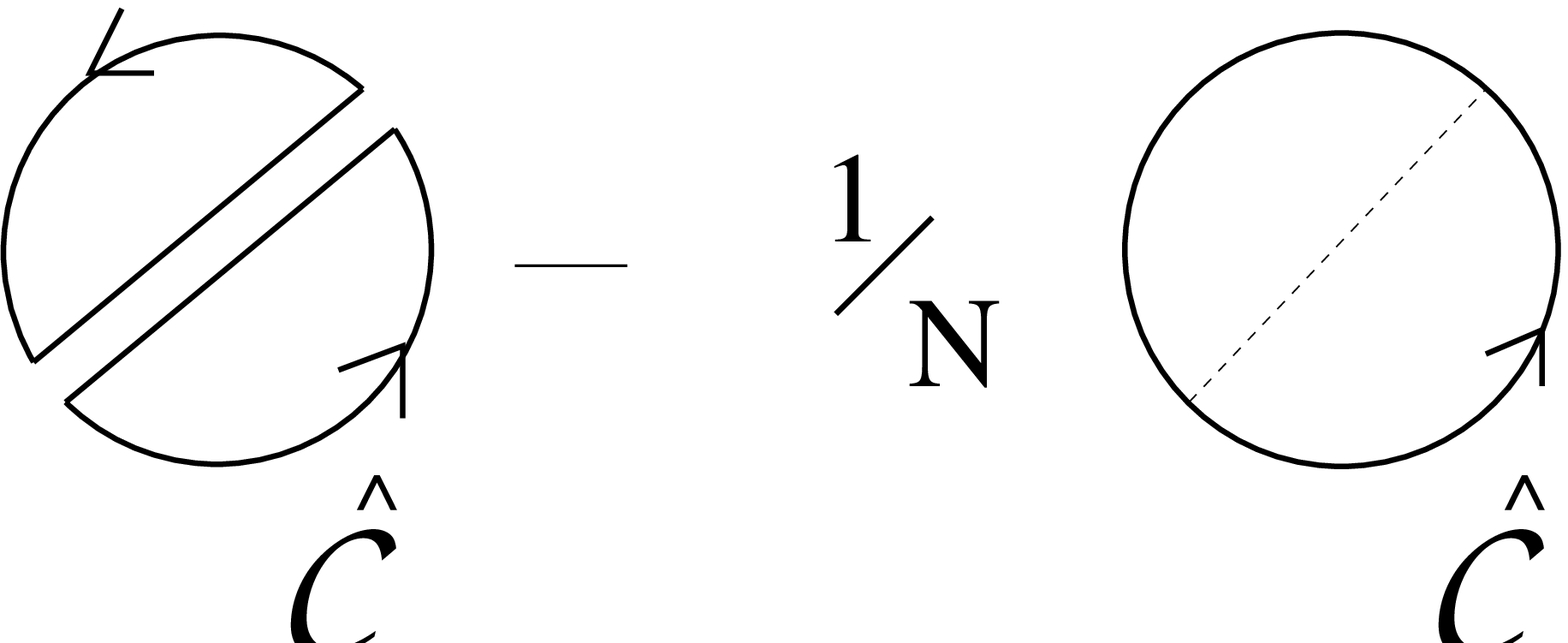}}\right\}
}$$
{\bf Fig.10.} The Wilson loop representation of the flow equation.
(The measure and associated functional derivative 
$\int\!\!\D\,\C\, {\delta/\delta\phi[\C]}$ are defined 
together, in the 
obvious way.\foot{See however the footnote in the Final Remarks.}) 
$$
\epsfxsize=.3\hsize\epsfbox{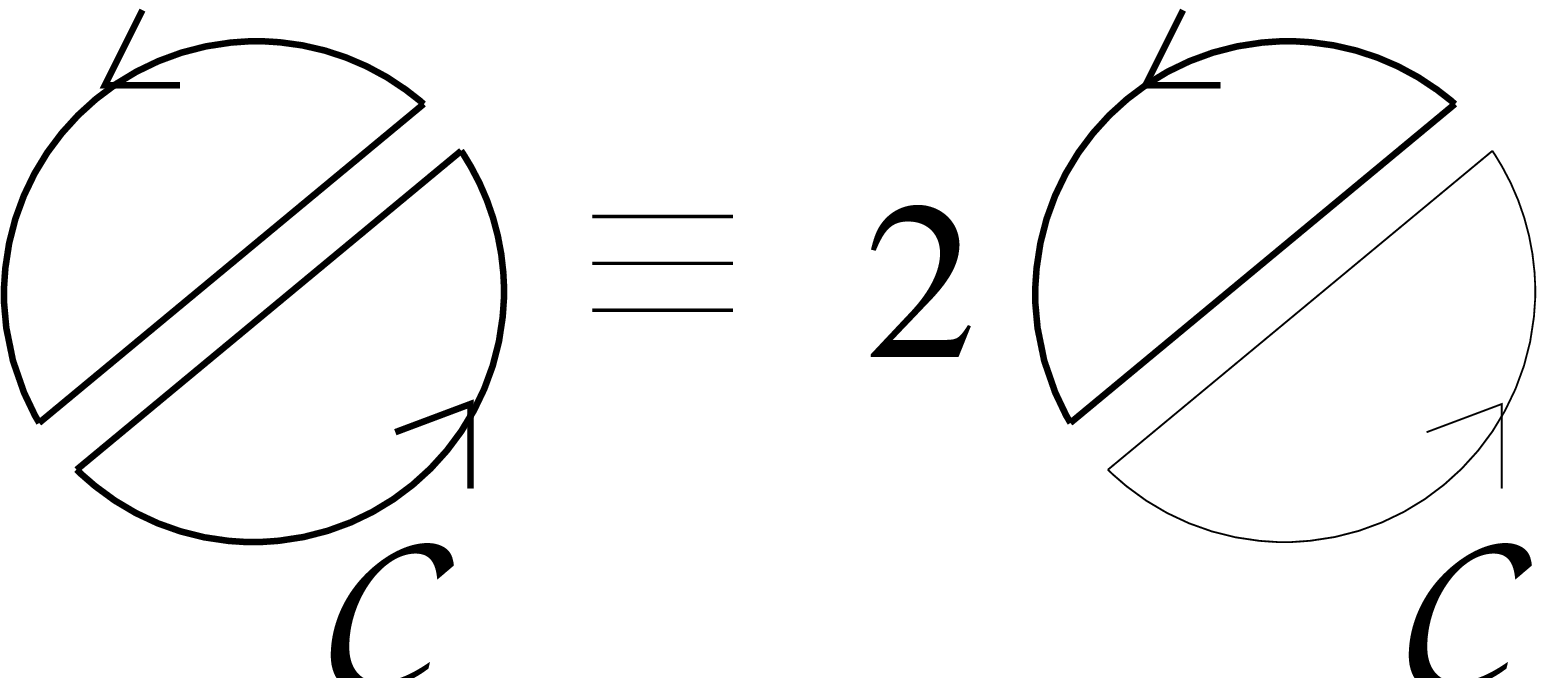}
$$
\centerline{ {\bf Fig.11.} The large $N$ limit;
the no-gauge-fields half contributes ${\rm tr} 1=N$.}

\newsec{The loop expansion}

Now, clearly the regularisation implied by the flow equation \erg, is
intimately related to covariant derivative regularisation. But we know
that such a regularisation fails at one loop. (It cures the superficial
divergences of all higher loops\oneslip.) Indeed, at one loop we get,
as usual, effectively the functional determinant of the kernel of the
bilinear term for small fluctuations $\A\mapsto\A+\delta\A$, \ie
\eqn\sone{S_1\sim\half\ln\D{\rm et}[-D^2c^{-1}(-D^2/\Lambda^2)]}
which in the UV limit $\sim\half\ln\D{\rm et}[(-D^2)^{r+1}]
\sim {r+1\over2}\ln\D{\rm et}[-D^2]$ clearly fairs no better than the
unregularised case! Since the problem now to be faced is proper regularisation,
particularly at one loop, we need to scrutinise the loop (\aka $\hbar$, or $g$)
expansion.

The solution to \erg\ may be expanded as
\eqn\Sloope{S={1\over g^2} S_0+S_1+g^2 S_2 +\cdots}
where $S_0$ is the classical (tree level) solution, $S_1$ the one loop
contribution, $S_2$ two loops and so on. It can be seen that the
$\beta$ function for $g$ must then take the standard form,
$$\Lambda{\partial g\over\partial\Lambda}=\beta_1g^3+\beta_2g^5+\cdots\quad.$$
Substituting these equations into \erg, we obtain
\eqnn\ergcl\eqnn\ergone\eqnn\ergtwo
$$\eqalignno{\Lambda{\partial\over\partial\Lambda}S_0 &=-a_0[S_0,S_0-2\hS]
&\ergcl\cr
\Lambda{\partial\over\partial\Lambda}S_1&=2\beta_1S_0-2a_0[S_0-\hS,S_1]
+a_1[S_0-2\hS] &\ergone\cr
\Lambda{\partial\over\partial\Lambda}S_2&=2\beta_2S_0-2a_0[S_0-\hS,S_2]
-a_0[S_1,S_1]+a_1[S_1]\quad, &\ergtwo\cr}$$
\etc The two-point solution of the classical flow equation \ergcl, is
$$S^0_{\mu\nu}(p)=2\Delta_{\mu\nu}(p)/f(p^2/\Lambda^2)$$
by gauge invariance and dimensions. We require $f(0)=1$ so
as to be consistent with \defg\ in the $g\to0$ limit. This fixes the solution
uniquely to be $f=c$, and thus 
\eqn\twop{S^0_{\mu\nu}=\hS_{\mu\nu}\quad.} 
Since by \defg, $S_{\mu\nu}(p)={2/ g^2}\,\Delta_{\mu\nu}(p)+O(p^3)$, we
have
$$S_{\mu\nu}(p)={1\over g^2}S^0_{\mu\nu}(p)+O(p^3)$$
which implies by \Sloope, {\sl that the $O(p^2)$ component of
all the higher loop contributions $S^n_{\mu\nu}(p)$,  must vanish}.
This determines the $\beta$-function. For example, in
solving \ergone\ for the one-loop two-point vertex $S^1_{\mu\nu}(p)$,
we have that the $a_0$ term vanishes by the equality \twop.
At $O(p^2)$, the l.h.s. of \ergone\ vanishes, while the
one loop contribution   $a_1[S_0-2\hS]$ ($\sim\Delta_{\mu\nu}(p)$
by gauge invariance) is generically non-zero. That leaves only
$2\beta_1S^0_{\mu\nu}(p)$ to balance this contribution, fixing
$\beta_1$. The higher $\beta_i$ follow similarly.

Most of our discussion has centred on Wilson loop pictures. We do not
want to give the impression that corresponding concrete realizations are
missing however! In fact these pictures translate directly into Feynman
diagrams. Thus for example, the classical two-point vertex satisfies
the diagrams in fig.12 (from \ergcl\ and the top two lines of fig.10).
$$
\Lambda{\partial\over\partial\Lambda}\,
\mathop{\vcenter{\epsfxsize=0.085\hsize\epsfbox{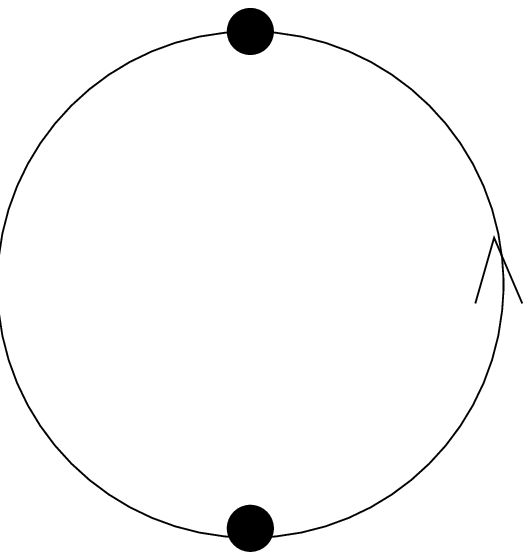}}^{\phantom{p}}
}^{\displaystyle
-p^\nu}_{\displaystyle p^\mu}=\ 
2\mathop{\vcenter{\epsfxsize=0.13\hsize\epsfbox{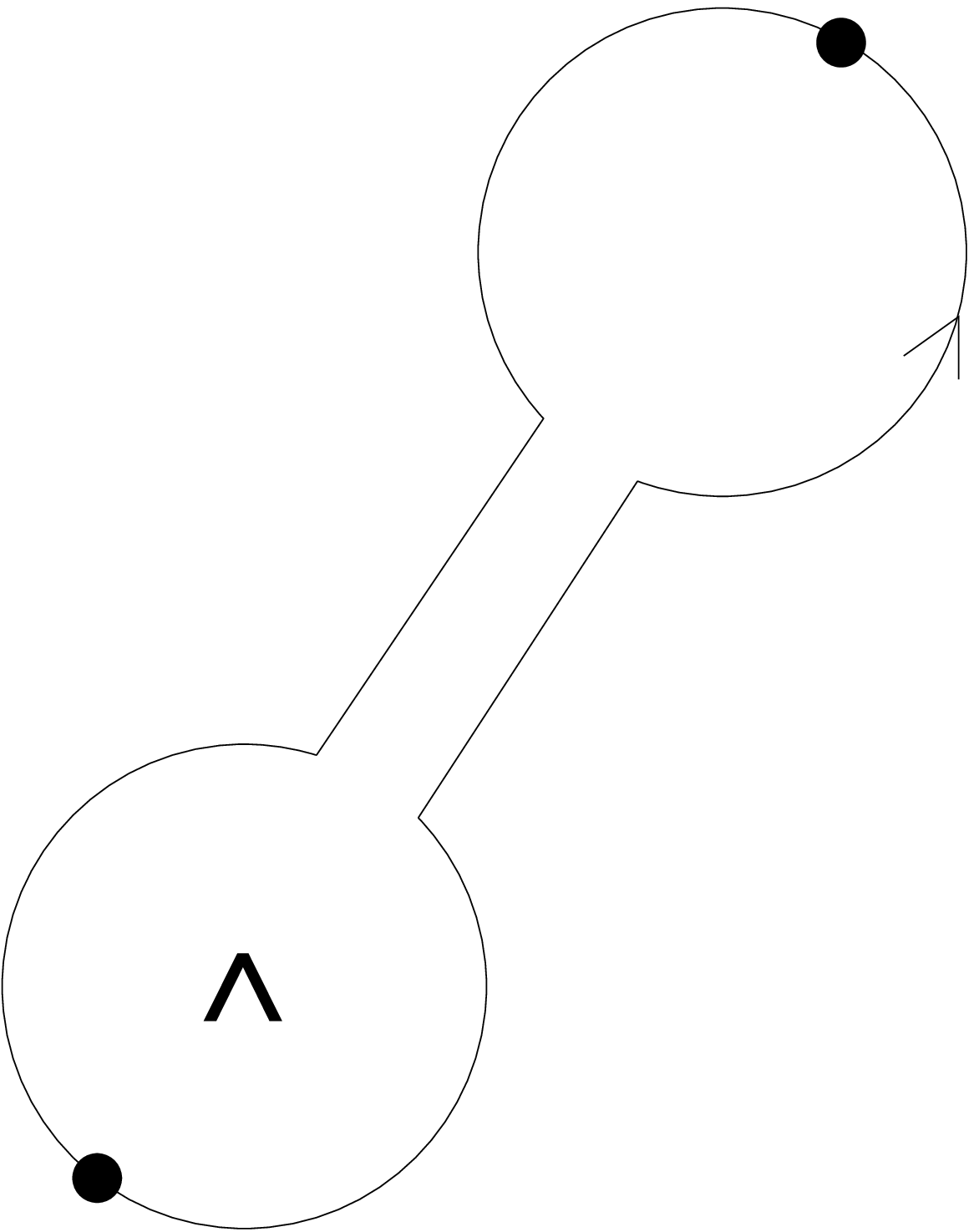}}}_{
\!\!\!\!\!\!\!\!\!\!\!\!\!\!\!\!\displaystyle p^\mu}
^{\displaystyle\;\;\;\;\;\;\;\;\;\;\;\;\;\;\;\;\;\;\;\;-p^\nu}
-\mathop{\vcenter{\epsfxsize=0.13\hsize\epsfbox{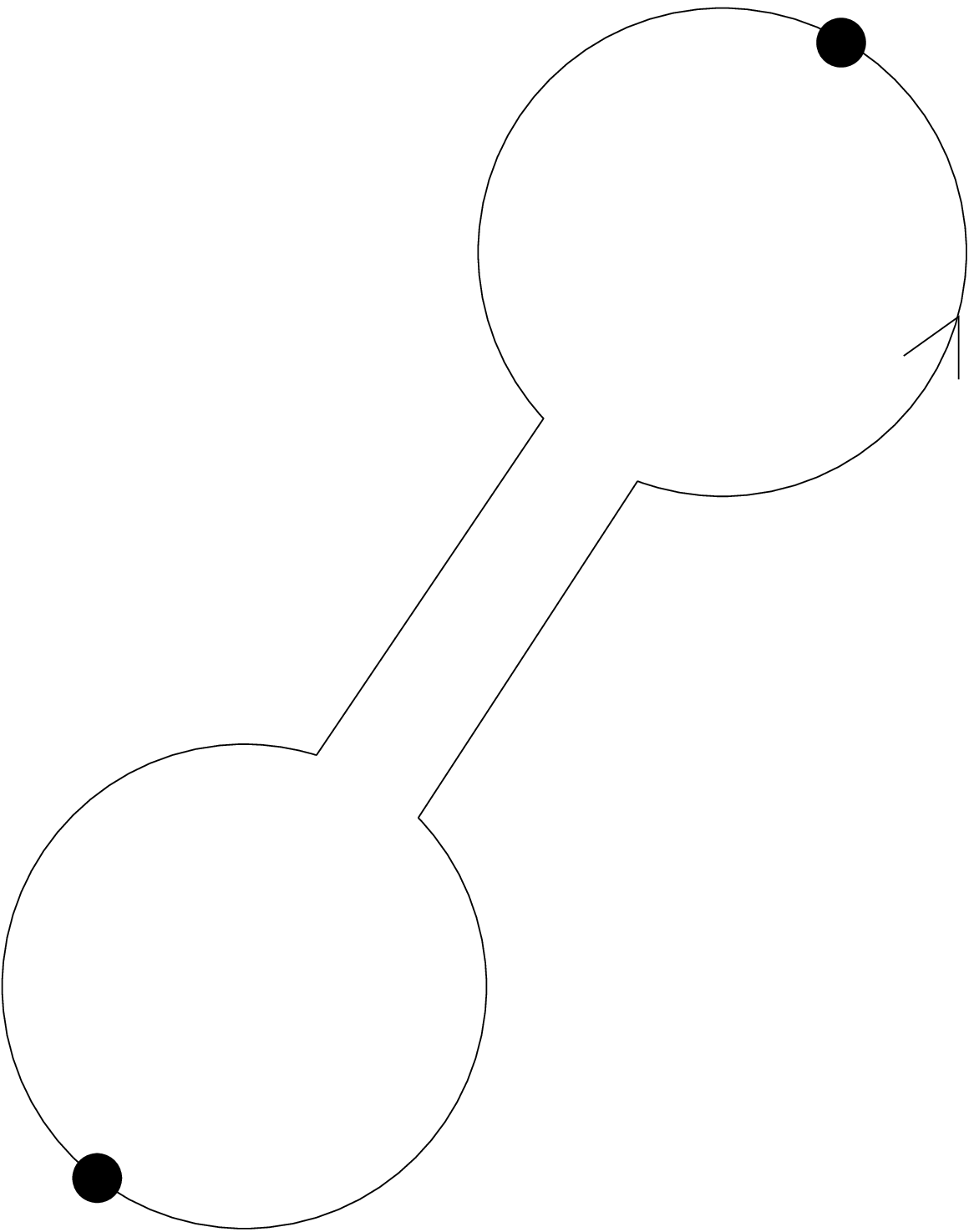}}}_{
\!\!\!\!\!\!\!\!\!\!\!\!\!\!\!\!\displaystyle p^\mu}
^{\displaystyle\;\;\;\;\;\;\;\;\;\;\;\;\;\;\;\;\;\;\;\;-p^\nu}
+(p^\mu\leftrightarrow -p^\nu)$$
\centerline{ {\bf Fig.12.} Feynman diagrams for the two-point vertex.}
Here we have put a caret in the Wilson loop that belongs to $\hS$.
The translation 
rule is just that we must place the points and their associated 
momenta in all places on the composite Wilson loop, 
while preserving the cyclic order. Since one-point
vertices vanish,\foot{for a whole host of reasons: Lorentz invariance plus
gauge invariance or momentum conservation, or charge conjugation (loop
reversal) invariance} we must have at least one blob per lobe. We see that
we just have a product of two two-point 
vertices connected by a wine, which since it has no gauge fields on it,
 is just given by $c'_p/(2\Lambda^2)$ from \defao\ and \aonorm. Thus, 
\eqn\dtwop{\Lambda{\partial\over\partial\Lambda}S^0_{\mu\nu}(p)=
{1\over2\Lambda^2}c'_p
\left[2\hS_{\mu\lambda}(p)-S^0_{\mu\lambda}(p)\right]
S^0_{\lambda\nu}(p)+(p_\mu\leftrightarrow -p_\nu)\quad,}
which, as already stated, is solved by \twop\ with \hSex.

At the three point level, we have the diagrams of fig.13. 
Here we have
used the relation \twop, which results in some diagrams cancelling
each other, and in particular ensures that
the three point vertex depends only on $n$-point seed 
vertices and already-solved lower-point vertices.
It is easy to see that
this feature holds for all higher point vertices.
$$
\Lambda{\partial\over\partial\Lambda}\
\vcenter{\epsfxsize=0.085\hsize\epsfbox{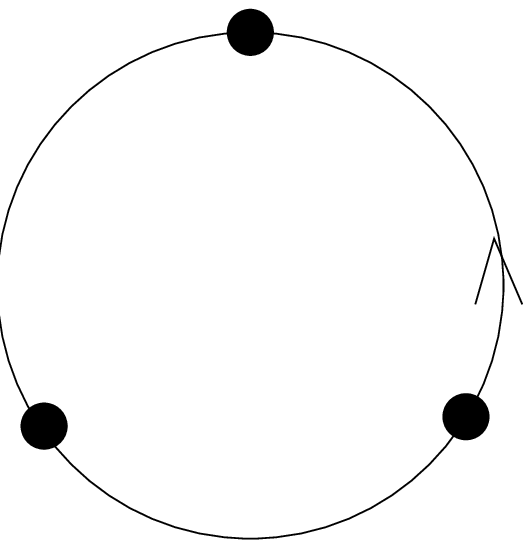}}\,
=\, 2\vcenter{\epsfxsize=0.13\hsize\epsfbox{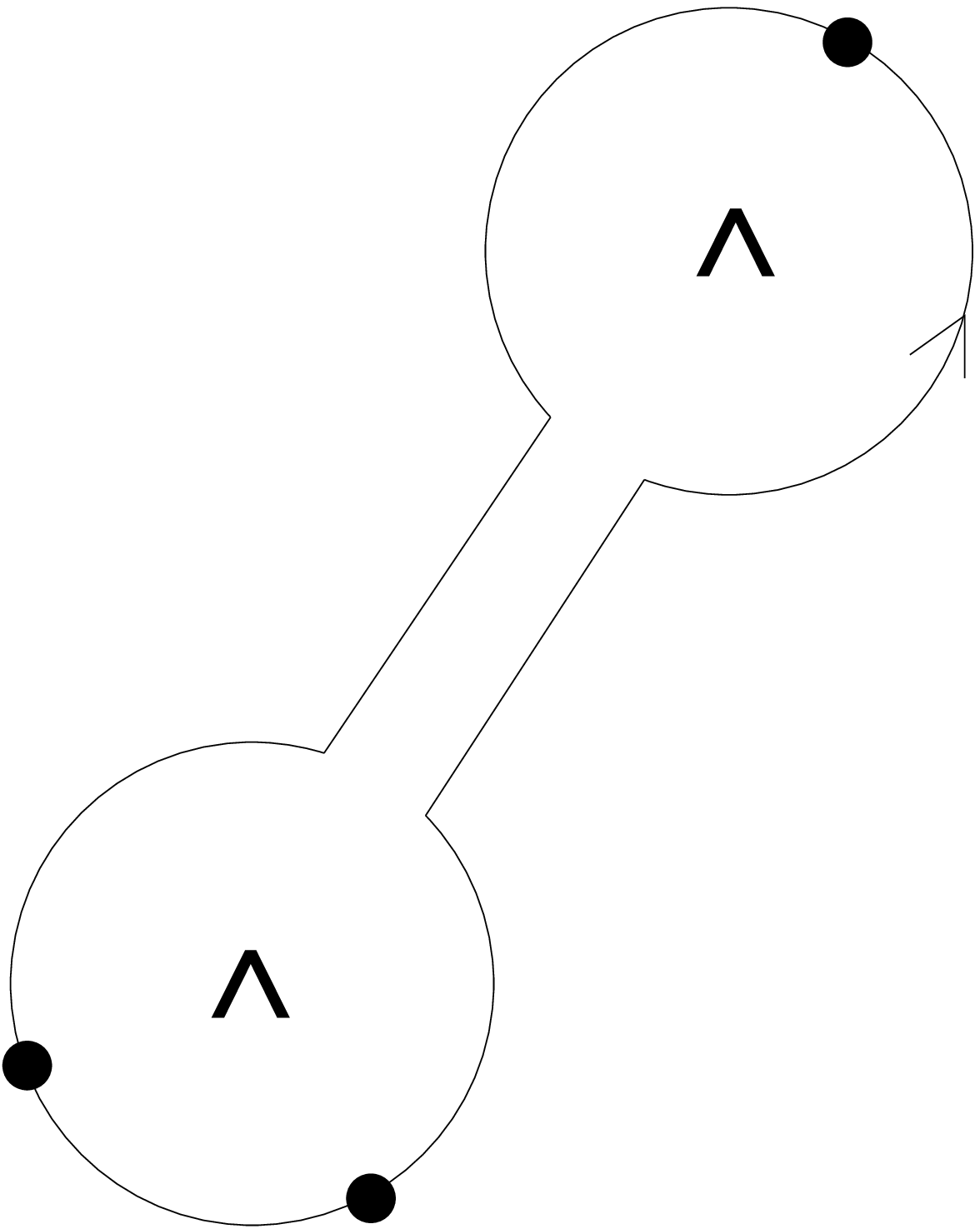}}
+2\vcenter{\epsfxsize=0.13\hsize\epsfbox{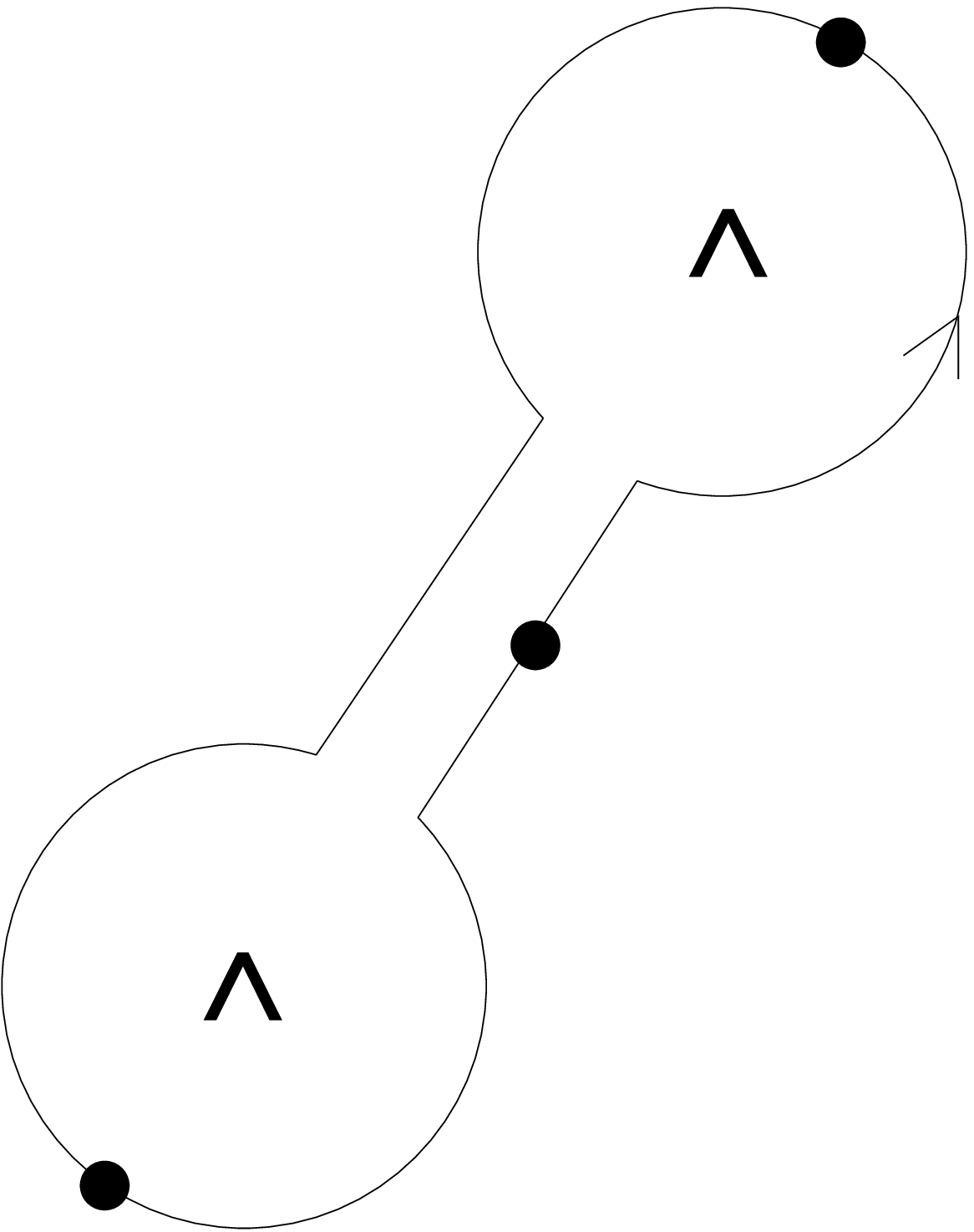}}
$$
{\bf Fig.13.} Feynman diagrams for the three-point vertex.
The r.h.s. should be summed over cyclic permutations of
the momentum labels.\hfill\break
 We can immediately integrate to give
\eqn\threep{\eqalign{
S^0_{\mu\nu\lambda}(\p,\q,\r)= &-\int_\Lambda^\infty\!\!{d\Lambda_1\over
\Lambda_1^3}\left\{c'_r\hS_{\mu\nu\alpha}(\p,\q,\r)\hS_{\alpha\lambda}(r)
+c'_\nu(\q;\p,\r)\hS_{\mu\alpha}(p)\hS_{\alpha\lambda}(r)
+{\rm cycles}\right\}\cr
&+2\Delta_{\mu\nu\lambda}(\p,\q,\r)\quad.}}
It should be understood that in the curly brackets we replace $\Lambda$
with $\Lambda_1$. The one-point wine is given by replacing $c^{-1}\mapsto
c'$ in \wone.
In principle the top limit would be $\Lambda_0$, the
bare cutoff scale, but by quasi-locality 
and dimensions the integrands have expansions
in (momenta)$^2/\Lambda_1^2$, so the continuum limit $\Lambda_0\to\infty$
exists. We only have the freedom to add an integration constant
$2\Delta_{\mu\nu\lambda}(\p,\q,\r)$, which can depend on momenta but not
on $\Lambda$. However even this freedom is illusory: we see from \threep\
that $2\Delta_{\mu\nu\lambda}$ is the three-point part of
$\lim_{\Lambda\to\infty}S^0$, which latter is fixed by \defg\
to be $-{1\over2}\int\!\!d^D\!x\, F_{\mu\nu}^2$.

Even though the formalism guarantees it, 
it is remarkable to see that
\threep, and the higher $n$-point solutions, explicitly yield manifestly
gauge invariant integrals over momentum scales: applying
a gauge transformation \ga, $\Delta_{\mu\nu\lambda}$ transforms into
differences of
the elementary two-point vertex $\Delta_{\mu\nu}$; the terms in
braces collapse to those on the r.h.s. of \dtwop, as is clear from figs.13,
7, and 12; the integral thus becomes exact, the top limit
cancelling the $\Delta_{\mu\nu}$ contributions and the bottom limit giving
the required differences of $S^0_{\mu\nu}$ -- the gauge transform of 
$S^0_{\mu\nu\lambda}$.

Now we address the question we have left hanging since the beginning of
the lecture. How do we know we are integrating out? In what
sense are higher momentum modes being suppressed? Unlike continuum
solutions of \pol\  for
example (\eg the
loop-wise solution for $\lambda\phi^4_4$ theory),
 here the three point and all higher point vertices 
{\sl diverge} in the limit of large momenta (
\ie large compared to $\Lambda$). 
In fact, given that the regularised two-point vertex diverges (as it must), 
the gauge invariance relations \ga\ {\sl insist}
that the higher point vertices diverge also. 
However by judicious choice of flow equation, as here,
these higher point vertices
diverge slower than the two-point vertex in the following sense:
For fixed momenta and $\Lambda\to0$, we
rescale the fields $A_\mu(\p)\mapsto A_\mu(\p)\sqrt{c_p}$ so as
to normalise the kinetic term (temporarily
mangling the gauge invariance). The three-point vertex has two
kinematic regimes: one controlled by the gauge relations \ga\
(which holds for example when all three momenta are colinear)
where $pS^0_{\mu\nu\lambda}\sim S^0_{\mu\nu}$, and 
a `transverse' regime which from \threep\ at least
diverges no worse than this. Thus the normalised vertex
$\sim S^0_{\mu\nu\lambda}c^{3/2}(p)\sim p \sqrt{c_p}$ vanishes as $\Lambda\to0$
[as do the normalised vertices of \pol]
showing that, as required, contributions of
modes of any fixed momentum (here attached to three-point vertices in a Feynman
diagram),
are suppressed completely in this limit.

Finally, we take a brief look at one loop. 
The diagrams for the two-point contribution
are displayed in fig.14.
$$
2\ \vcenter{\epsfxsize=0.085\hsize\epsfbox{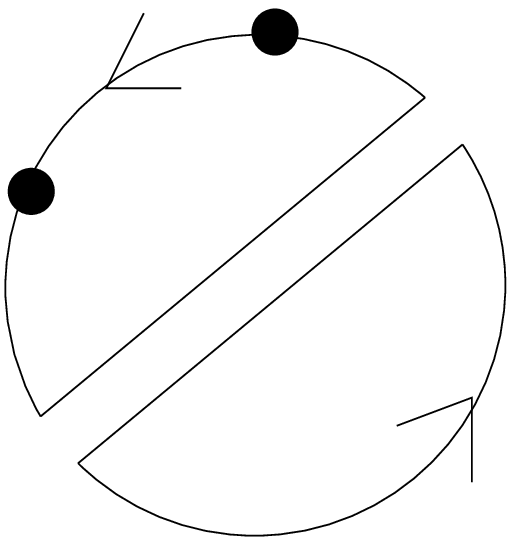}}
+2\ \vcenter{\epsfxsize=0.085\hsize\epsfbox{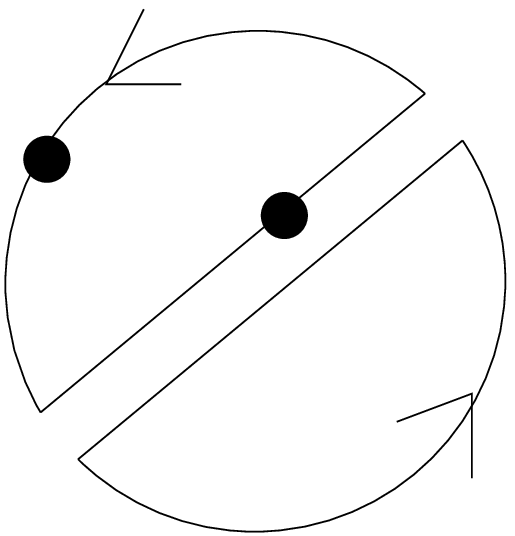}}
+2\ \vcenter{\epsfxsize=0.085\hsize\epsfbox{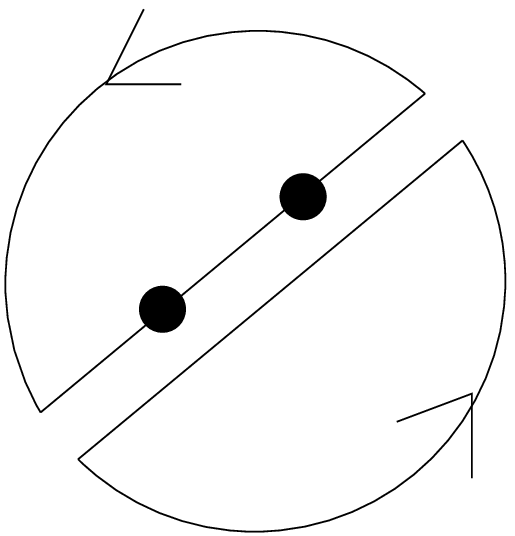}}
-{2\over N}\ \vcenter{\epsfxsize=0.085\hsize\epsfbox{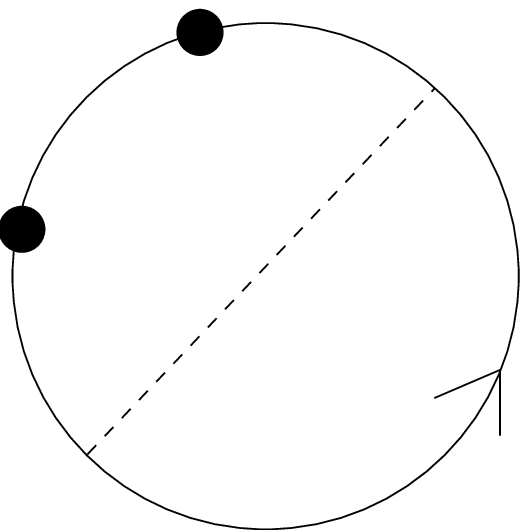}}
-{1\over N}\ \vcenter{\epsfxsize=0.085\hsize\epsfbox{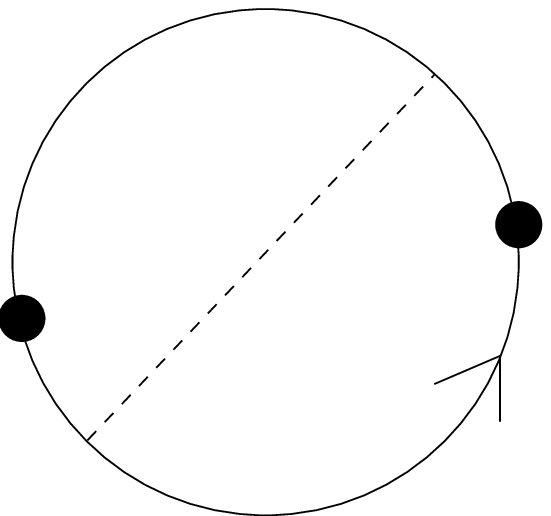}}
$$
\centerline{ 
{\bf Fig.14.} The one-loop two-point diagrams. The
Wilson loop stands for
$\Sigma=2\hS-S_0$. }
The $1/N$ contributions cancel each other and the two 
$p_\mu\leftrightarrow-p_\nu$ permutations combine, 
after use of symmetries and flow equations.
Translating the Feynman rules, we thus have
\eqn\ol{\eqalign{
2{N\over\Lambda^2}\int\!{d^Dk\over(2\pi)^D}\Big\{
&c'_k\Sigma_{\alpha\mu\nu\alpha}(\k,\p,-\p,-\k)
+c'_\mu(\p;\k-\p,-\k)\Sigma_{\alpha\nu\alpha}(\k,-\p,-\k+\p)\cr
+&\Sigma_{\alpha\alpha}(k)c'_{\mu\nu}(\p,-\p;\k,-\k)\Big\}\quad,} }
and recalling the
discussion on $\beta_1$, this must equal
$4\beta_1\Delta_{\mu\nu}(p)+O(p^3)$.
We know \ol\ is transverse by gauge invariance. Indeed contracting
with $p_\mu p_\nu$, it is straightforward to show that the above
collapses to
\eqn\gilch{2{N\over\Lambda^2}\int\!{d^Dk\over(2\pi)^D}\left\{ c'_{\k+\p}
\hS_{\alpha\alpha}(\k+\p)-c'_k\hS_{\alpha\alpha}(k)\right\}\quad,}
which of course vanishes after shifting the momenta.

As it stands however, the $k$ integral in \ol\ is,
in $D=4$ dimensions, quadratically divergent.
(The same is true for the higher-point vertices.)
You might think that gauge invariance plus dimensions implies that
the divergence would be at worst logarithmic, but the coefficients
are {\sl not} constrained by such `power counting': we obtain 
many extra powers of 
momenta balanced by factors of $1/\Lambda$.
This makes life complicated. In fig.15 we show two examples of
one-loop contributions, where we have expanded the composite Wilson loops
in fig.14. In the first example the top lobe is the classical three-point
function (expanding to fig.13). The dependence of the divergence 
on $p$ is thus not even
polynomial. ({\it N.B.} Actually, for this contribution not to vanish for
Lorentz symmetry reasons, we have to include an extra point on the bottom lobe
or in the loop. The divergence is then quadratic.)
In the second example, we see that the gauge invariance
restriction to transverse $\sim\Delta_{\mu\nu}(p)$, has no effect on
the UV divergence $\sim \hS_{\mu\alpha}(p)\hS_{\alpha\nu}(p)
\int\!{d^D\!k/ k^2}$. 
$$
\epsfxsize=.6\hsize\epsfbox{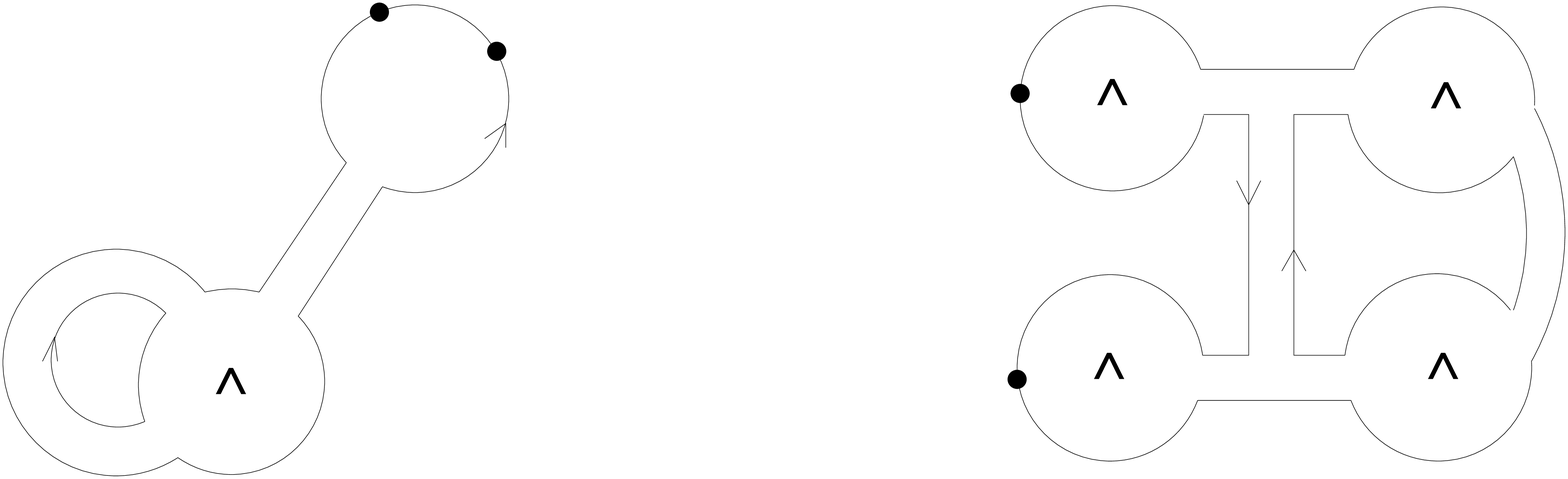}
$$
\centerline{ {\bf Fig.15.} Two divergent one-loop contributions in \ol.}

We can try to avoid our conclusions 
about the one-loop failure of higher derivatives \sone, by
carefully balancing new higher-point interactions (without modifying
the rank of $c^{-1}$), which in this context can be introduced 
ultralocally in $\hS$ or quasilocally in the wine. Since the
worst divergence is quadratic, its coefficient must be gauge invariant.
If this is cancelled by these modifications, then the worst divergence
is logarithmic with a gauge invariant coefficient, which we again
seek to cancel by appropriate modifications. In fact, these requirements
of locality leave us very little choice over the modifications and 
following this route we succeeded in ensuring one-loop finiteness of the 
two-point vertex to all orders in $p$. The problem is that this 
makes matters worse for the three and higher point vertices, and in
any case convergence at higher loops is unclear. It could well be that
there is no solution along these lines.
At any rate we were persuaded to embark on a
search, using Pauli-Villars
ideas, for a simpler, more natural, regularisation.
This will be described in the third and final lecture.

\newsec{Pauli-Villars regularisation}

The basic idea  is to introduce an
action bilinear in some PV fields $S_B\sim\int \Bb{\cal P}B$
so that on integrating them out, the partition function is modified to
\eqn\Zpv{\Z\sim\int\!\!\D A\ \D{\rm et}^{j/2}{\cal P}\ \e{-S}\quad,}
where $j$ is the (signed) number of  PV fields (minus for bosons).
The kernel ${\cal P}$ is chosen so that the resulting one-loop divergences
cancel those left by the covariant higher derivative regularisation \pv\warr.
As is well known, a one-loop contribution $\sim\ln\D{\rm et}{\cal P}$ can
(again) be represented as a path integral for a particle circulating 
in a loop. But we cannot simply redefine the effective action
$S\mapsto S-{j\over2}\ln\D{\rm et}{\cal P}$,
because the second term is divergent.
If it is really true (as it seems)
that effective gauge theory cannot be regularised without
such PV contributions, then this says something important:
exact gauge invariance at the effective level requires the
existence of a measure term which is not separately
well-defined (only the kernel ${\cal P}$ being well-defined).

This idea of bilinear PV fields however is not so neat in 
practice\foot{complications arise from higher-loop
`overlapping' divergences}
\pv\warr, nor can the property of bilinearity be preserved in this context,
as illustrated in fig.16. 
$$
\epsfxsize=.18\hsize\epsfbox{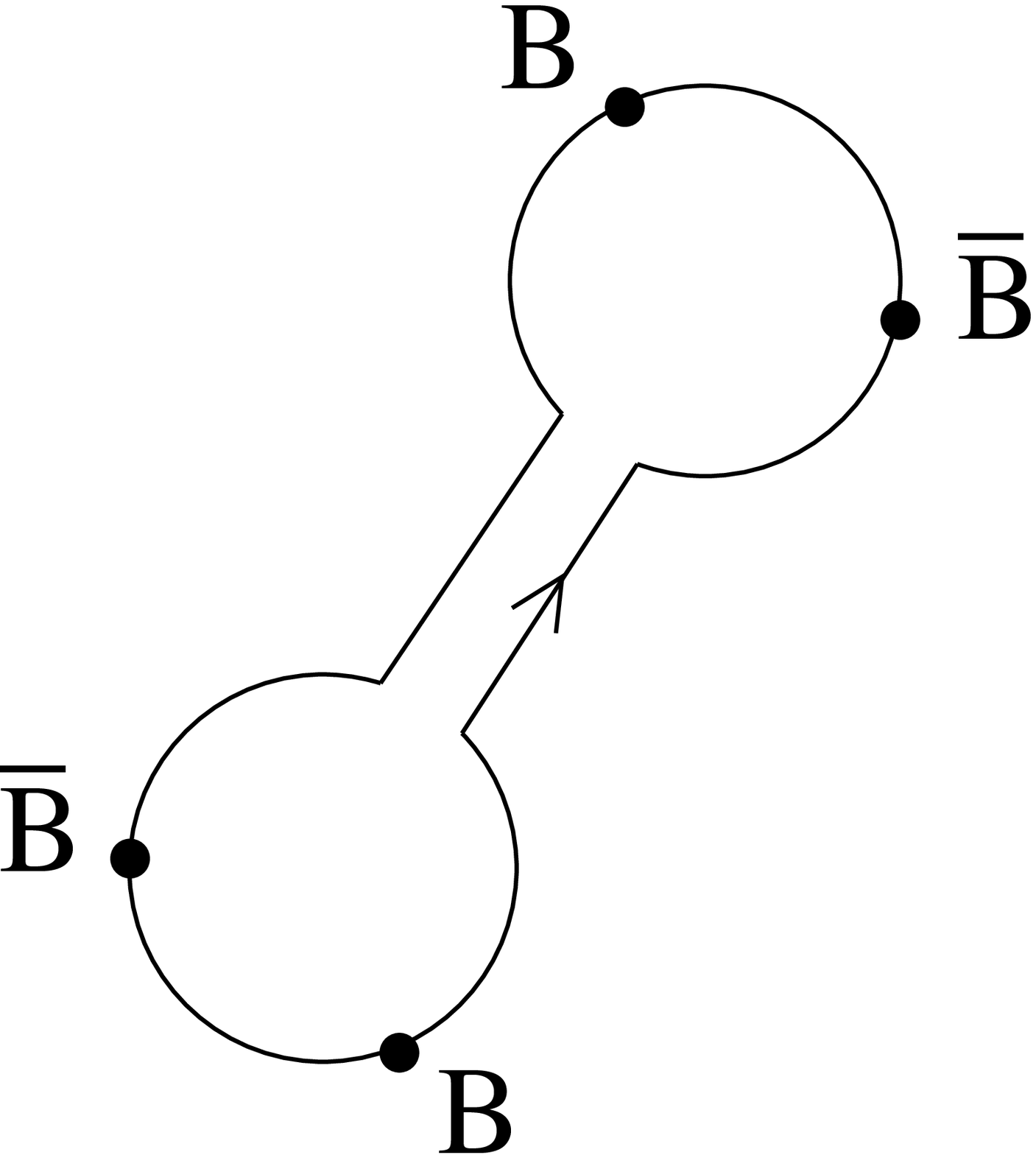}
$$
\centerline{ {\bf Fig.16.} A four-point PV interaction generated on 
integrating out the gauge fields.}
Note especially that the extra interactions
generated in this way by the flow, generically will not properly
regularise the theory at the effective scales
$\Lambda<\Lambda_0$.
We must face up to the problem of
multipoint PV interactions with, as we have seen, little help
from power counting arguments.

Working purely on a heuristic level, let us recall that Pauli-Villars
regularisation usually works as follows. We want to replace the propagator,
which for the gauge fields is $\sim(\Delta_{\mu\nu}/c)^{-1}$,
by a subtracted propagator
\eqn\subi{\sim\ \left(\Delta_{\mu\nu}/c\right)^{-1}-
\left(\Delta_{\mu\nu}/c+\Lambda^2\delta_{\mu\nu}\right)^{-1}\quad.}
The idea is that in the UV limit the two propagators cancel, 
making UV finiteness obvious, while
at energies much lower than the cutoff $\Lambda$, only the original
propagator survives. The second propagator will be provided by
vector PV fields. To cancel
the one loop determinant \sone\ from $A_\mu$, they must be fermions
and to avoid being symmetrized away, complex. But this is not
the end of the story, because the original propagator lives only in
the transverse space whereas the second propagator has a 
point-like longitudinal
contribution $\delta_{\mu\nu}/\Lambda^2$ with bad UV properties.
Therefore we regularise this term with higher derivatives:
\eqn\kB{S_B\sim{\rm tr}\int \Bb_\mu\left(\Delta_{\mu\nu}/c
+\Lambda^2\delta_{\mu\nu}/\ct\right)B_\nu}
($\ct(0)>0$), where $\ct^{-1}$ is a polynomial in $p^2/\Lambda^2$
of rank ${\tilde r}$, with
${\tilde r}<r+1$ so as not to disturb the cancellation
properties of \subi. But covariantized, these higher derivatives will 
once again not succeed just as $c^{-1}$ failed in the transverse
sector.
Therefore we need to subtract from the 
longitudinal part of the $\Bb$--$B$ propagator another 
purely longitudinal term, which can be generated by 
\eqn\kC{S_C\sim\half{\rm tr}\int\ D_\mu C\,(\Lambda^2/\ct)\,D_\mu C 
+\sigma\Lambda^4C^2\quad,}
where $C$ is a boson and for the same reasons as before we give it
a mass of order the cutoff. Now the propagators are well regulated
and effectively appear as
\eqn\modprop{\sim \left({\Delta_{\mu\nu}\over c} \right)^{-1}-
\left({\Delta_{\mu\nu}\over c}+\Lambda^2{\delta_{\mu\nu}\over\ct}
\right)^{-1}+p^2\delta_{\mu\nu}\left({\Lambda^2\over\ct}p^2
+\Lambda^4\sigma\right)^{-1}\quad,}
where the $p^2\delta_{\mu\nu}$ numerator on the $C$--$C$ contribution
appears through its derivative interactions. 
$$
\epsfxsize=.7\hsize\epsfbox{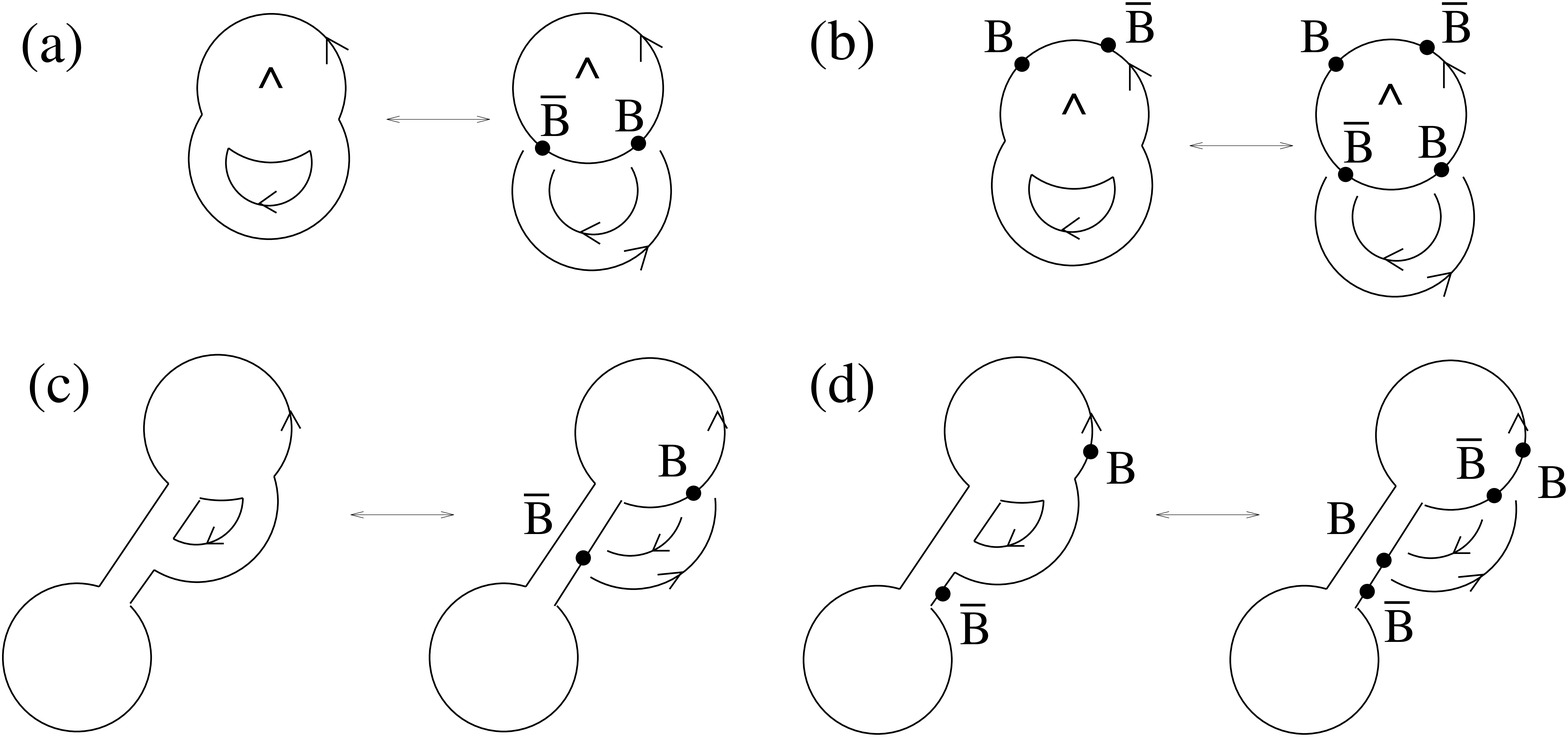}
$$
\centerline{ {\bf Fig.17.} Building PV regularised `propagators' via
pair-wise cancellations.}
Now, in order for such a regularisation to work we need at least one
`propagator' (actually wine)
in every loop to match up as \modprop. It is easy to see
that this implies that the seed action and the wines must be `peppered'
with $\Bb$ $B$ pairs to all orders. Thus 
consider loops formed from a wine and $\hS$. 
Working to all orders in $A$, we see, as in fig.17a,
that we must have pair-wise cancellation with a
$\Bb$ $B$ vertex, constructed from $\hS$ 
by replacing two $A$'s by a $\Bb$ $B$ pair. This implies, since the
wine can catch any gauge field, that there better be
 a similarly constructed
$(\Bb\ B)^2$ vertex  as we see in fig.17b, and so on; 
similarly for the wines (figs.17c and d). Note that the order, with
$B$ always following $\Bb$, is forced on us \eg we cannot have 
$B$ $B$ vertices constructed in this way from $\hS$,
because they vanish by conflict of Bose/Fermi symmetries. 

There are problems with this approach however. 
Since the wine can bite anywhere, we will get $B$s following $B$s
round corners where a wine joins a Wilson loop (ditto $\Bb$s).
This is a problem because the $1\leftrightarrow1$ matching will
then fail: we can make more loops with $\Bb$--$B$s in, than 
equivalent $A$ loops. It is also in conflict with our 
Wilson loop picture: a pair of adjacent $B$s mark a corner, 
which makes no
sense if they are merely `decorating' an underlying fluctuating loop.
Finally, the effective degrees of freedom 
of $A$, $B$ and $C$, which are
$D-1$, $-2D$ and $1$ respectively, all $\times(N^2-1)$, do not match.
Here we have noted that the gauge field only communicates through
its transverse degrees of freedom [\cf \modprop], 
while $B$ has no such constraint, is complex and with
wrong sign statistics. We would have expected the degrees of
freedom to cancel, reflecting our required almost-perfect
cancellation at high energies \modprop.

All these problems are solved by the following idea. We 
take the hint from the counting argument, and double the
gauge group to $SU(N)\times SU(N)$. The original gauge and longitudinal
PV field will be written $A_\mu^1$ and $C^1$. We introduce  $A^2_\mu$ 
and $C^2$ belonging to the second $SU(N)$. We place 
$B^{i_1}_{j_2}$ in the `middle', 
fundamental with respect to $SU_1(N)$ and complex conjugate fundamental
in $SU_2(N)$ (thus oppositely for $\Bb^{i_2}_{j_1}$). This way $B$ 
{\sl must} follow $\Bb$ and \vv, by group theory \cf fig.18. 
$$
\epsfxsize=.33\hsize\epsfbox{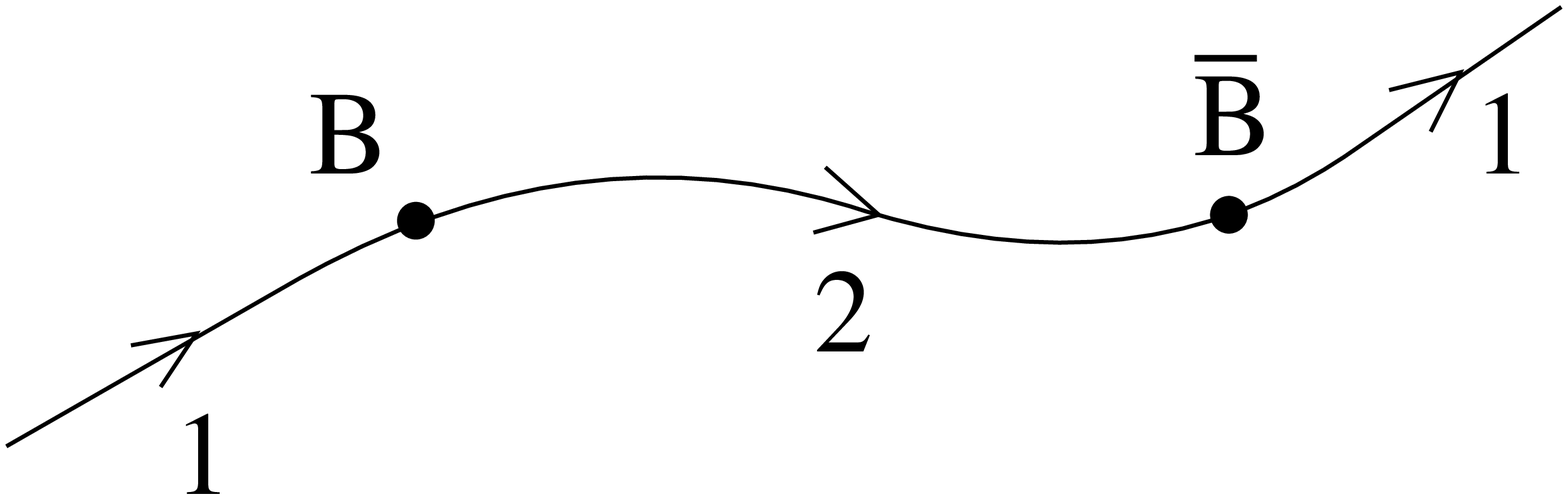}
$$
\centerline{ {\bf Fig.18.} Group theory forces $\Bb$s to follow $B$s.}
The degrees of freedom  
of $A^i_\mu$, $B_\mu$ and $C^i$, \viz $2(N^2-1)(D-1)$,
$-2N^2D$ and $2(N^2-1)$ respectively, now cancel as expected
-- at leading order
in large $N$. At finite $N$ the counting argument suggests that
we need some group-singlet Lorentz-vector PV bosons. It makes sense 
however to solve first the $N=\infty$ case, and in fact
from here on we will describe only this case.

Now everything falls into place. In the large $N$ limit every
loop appears in two flavours,  whose UV limits
should pairwise cancel \cf \eg fig.19: by group theory,
the wine in the first diagram
of fig.19 propagates $A^1$ or $C^1$, while the wine in the second must be
$\Bb$--$B$. 
$$
\epsfxsize=.35\hsize\epsfbox{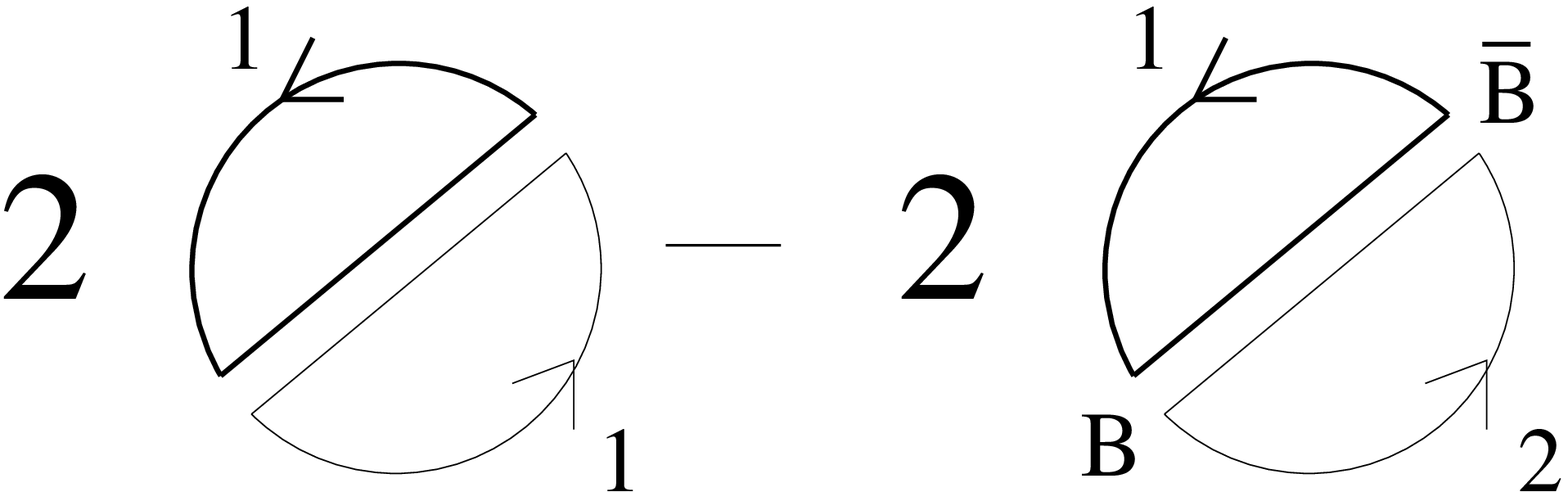}
$$
\centerline{ 
{\bf Fig.19.} Pairwise cancellation in the large $N$ limit.}
(For clarity, we mark the position of the $B$ and
$\Bb$ attachments.)
This in turn suggests
that `2'-type Wilson loops count negatively, which in turn would
imply a wrong sign $A^2$ action. Actually this has to be the case
if pairwise cancellation is to work, as we now show.

Since the two diagrams in fig.20a must pairwise cancel (in the UV
limit), the two wines in fig.20b must have opposite signs.\foot{
Signs are referred to a `canonical'
Fermi ordering, \eg
$\Bb$ before $B$, $B$ before ${\delta/
\delta B}$, \etc}
 The two diagrams in fig.20c
are both positive because they each contain two actions and two wines
of the same flavour.
Since each of these two diagrams must be cancelled pairwise by attaching a
$\Bb$--$B$ wine respectively to each of the two diagrams in fig.20d, 
we conclude that the pair in fig.20d must have the same sign. But
since the $A^1$ action in fig.20d must be positive, and by fig.20b,
the two wines have opposite signs, this implies that the $A^2$
action is negative.
$$
\epsfxsize=.75\hsize\epsfbox{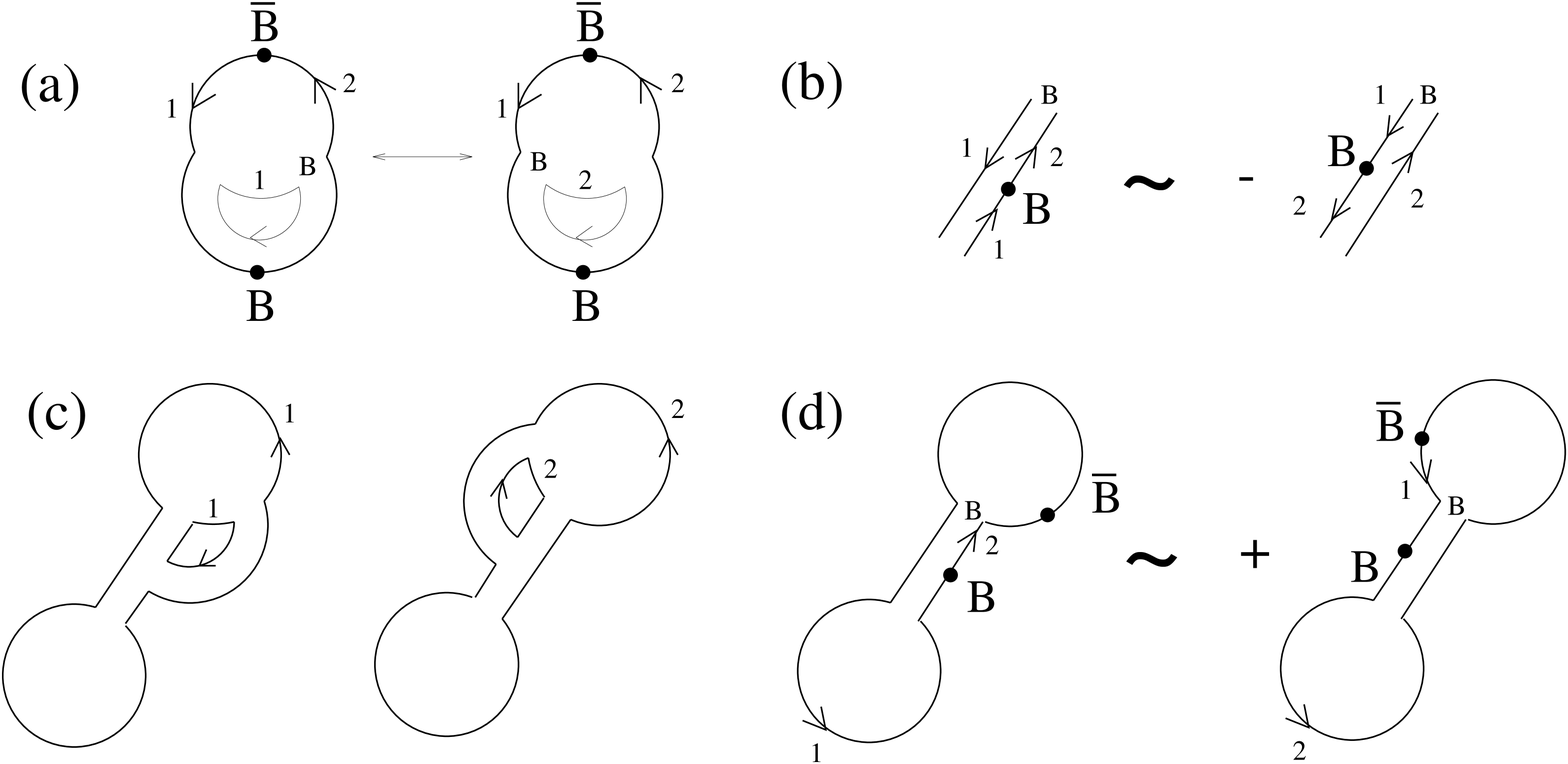}
$$
\centerline{ {\bf Fig.20.} Demonstrating that $S^{(2)}$ must have
 the wrong sign.}

The wrong sign $\hS^{(2)}$ (and corresponding $A^2$ wine)
is not as bad as it might seem at first sight.
$A^1$ and $A^2$
only talk to each other through the massive $B$ fields,
thus at energies much less than $\Lambda$ we obtain two 
decoupled pure gauge theories -- the physical $SU_1(N)$ and an
unphysical $SU_2(N)$. Furthermore, it can be shown that
these changes of sign in the `2' sector
are equivalent as far as this sector is concerned,
to changing the sign of $g^2$. Thus even in the `2' sector
the loop expansion to all orders is well-defined. Therefore
if any instability is communicated to the $A^1$ sector,
it can do so only at the \nonp\ level.

The fact that the diagrams of fig.20d both have the same sign, illustrates
an important general consequence: the sign of a 
composite Wilson loop cannot
depend on where around the loop the cyclically ordered $B$s and $\Bb$s
are placed. Indeed if the sign were to change on moving, say, a $B$
from a wine to an adjacent lobe, then one of the two diagrams cannot
cancel its partner. Again, it also has to be true if we take seriously
the underlying one-particle picture. 

By `injecting' extra $\Bb$ $B$ pairs,
as in fig.21, and dragging them into the rest of the loop we can thus 
determine the signs of all higher point  $\Bb$ $B$ interactions.
$$
\epsfxsize=.2\hsize\epsfbox{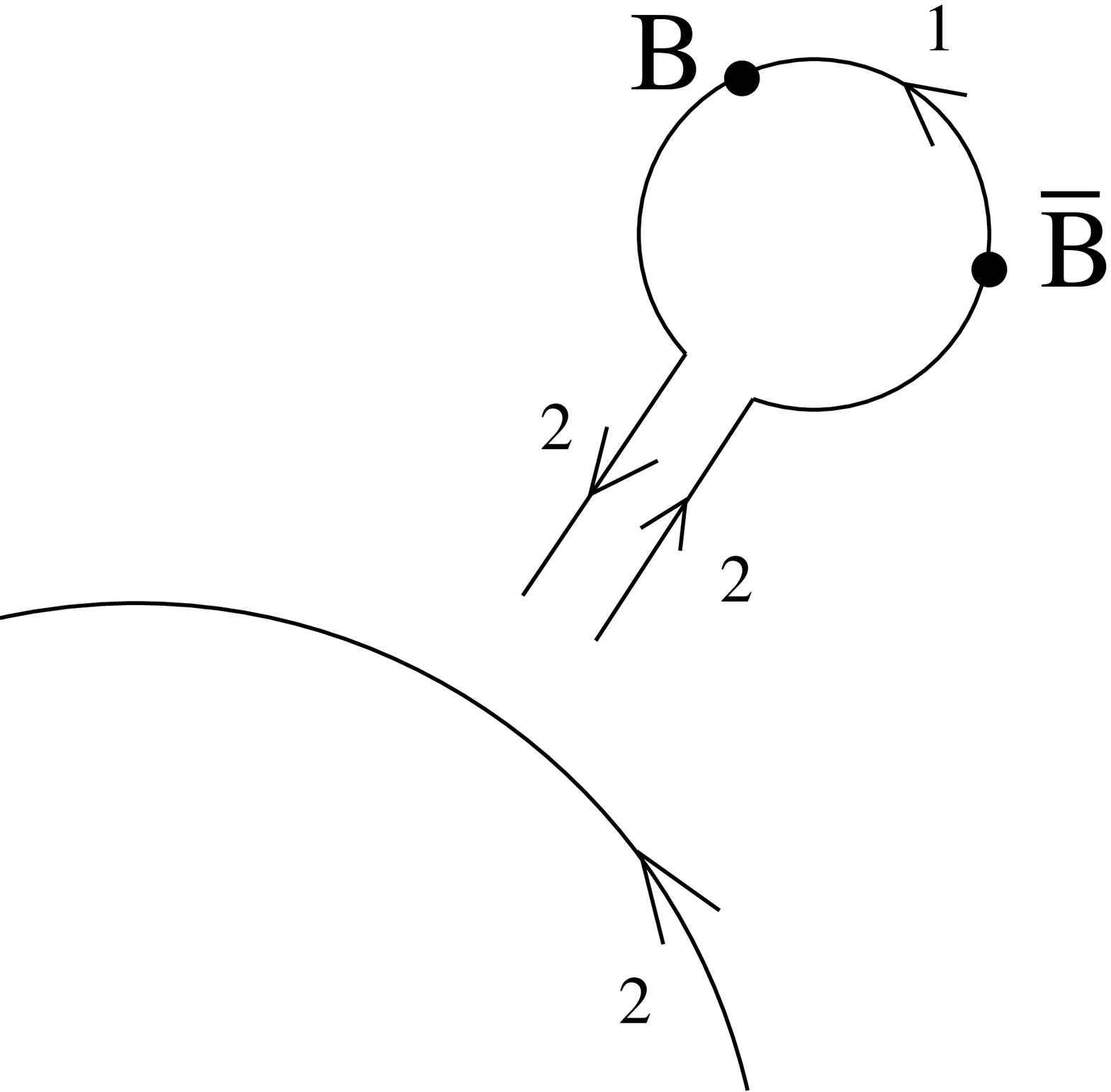}
$$
\centerline{ {\bf Fig.21.} `Injecting' a $\Bb$ $B$ pair.}
The requirement of pairwise 
cancellations fixes uniquely all the
 signs (up to global changes of sign via
$B\mapsto-B$, or $\Bb\mapsto-\Bb$), and the result
can be expressed \nonp ly in the $B$ fields, by introducing the `peppered'
Wilson line
$$\Phi[\C]= P\exp\left\{\int_{\tau_1<\tau_2}\!\!\!\!\!\!\!\!\!\!\!
d\tau_1d\tau_2\, 
{\dot z}\!\cdot\!\Bb(\tau_1)\, {\dot z}\!\cdot\!B(\tau_2)\ 
+\int\!\!d\tau\, {\dot z}\cdot A \right\}\quad,$$
where the path ordering is done with respect $A$ and non-overlapping pairs
of $\Bb$ $B$, and the flavour $A^i$ should be chosen according to group
theory, \ie as in fig.18.

The normalisations of the wines [\cf \aonorm] are fixed by the requirement
that the kinetic terms \kB, \kC\ are reproduced in $\hS$ and $S^0$
[via the analogue
of \ergcl]. Thus it turns out that the $\Bb$--$B$ wine covariantizes 
 $K^p_{\mu\nu}$, where
$K^p_{\mu\nu}=\delta_{\mu\nu}K_p+{p_\mu p_\nu\over\Lambda^2}L_p$, with
$$K(x)={d\over dx}\left({x\ct c\over x\ct+c}\right)\ins11{and} 
xL(x)={d\over dx}\left({x^2\ct^2\over x\ct+c}\right)\quad,$$
while the $C$--$C$ wine covariantizes $M_p$,  where 
$$xM(x)={d\over dx}\left({x^2\ct\over x+\sigma\ct}\right)\quad.$$
These have the nice cancellation properties we required, with $K\sim c'$
and $L\sim M\sim (x\ct)'/x$ at large $x$. Actually, this does not fix the normalisation
of the `mixed' wines such as those in fig.20b. They get fixed by their
interaction with the $C$ sector.

Recall from \modprop, that the $C$ terms are needed to UV
cancel the longitudinal part from $\Bb$--$B$. We almost
do not need them because we can add the longitudinal 
contribution directly into the $A$--$A$ wine, augmenting
\eqn\AAL{c'\delta_{\mu\nu}\mapsto c'_{\mu\nu} := c'\delta_{\mu\nu}
+{p_\mu p_\nu\over\Lambda^2}L\quad.}
This way the longitudinal cancellation is even exact, not just in the UV
limit. The addition of this term does not affect the gauge field
contributions, \eg the ones considered in the first lecture, because on
covariantizing, the longitudinal wine has at each end 
$D_\mu\cdot{\delta/\delta A_\mu}$ which, acting
on any pure gauge part of the action,
vanishes, by gauge
invariance. More generally,
such a longitudinal wine drifts to the nearest obstruction, as illustrated
in fig.22. 
$$
\epsfxsize=.55\hsize\epsfbox{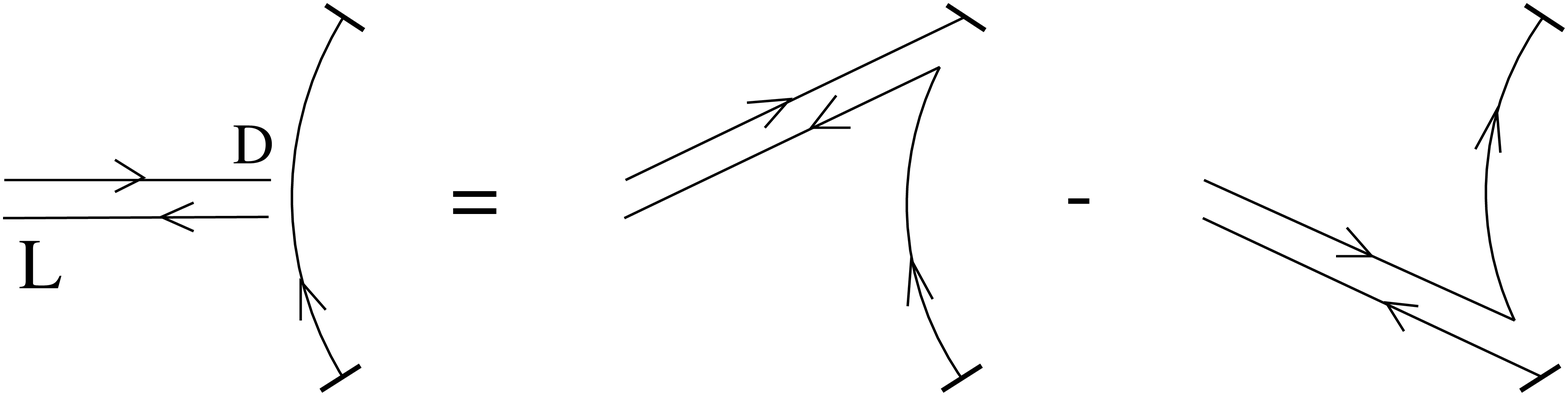}
$$
\centerline{ {\bf Fig.22.} A drifting longitudinal wine. Note the loss of
$D$ on the r.h.s. (\cf also fig.7.)}

You might think that peppered $B$s and $\Bb$s count as obstructions
but this is not the case. After all interactions are properly peppered,
the longitudinal wine sails through $B$ fields in the same way that
it sails through gauge fields. This is an important ingredient: 
many contributions
with longitudinal wines would otherwise
fail to converge. Consider for example that of fig.23.
This has a contribution
where the left lobe is constructed from peppering $\hS^{(1)}$, and whose UV
behaviour goes as $c^{-1}_k/k^2$, and where the right lobe comes from the
$\ct^{-1}$ part of \kB. By fig.22, $L$'s attachment to the left lobe 
drifts up to the
$K$ wine and down to $\Bb$. The first case is finite,
because 
a vertex with touching wines diverges at worst linearly.
The second case leads na\"\i vely
to an (in $D=4$ dimensions) overall cubic divergence: $c^{-1}_k/k^2
 c'_k\ct^{-1}_k kL_k \sim O(1/k)$. Furthermore, there is no 
pairwise cancelling partner, because there is no
$A$ analogue of the $\ct^{-1}$ part of \kB. 
Fortunately after taking into account the other
pepperings of fig.23, the $\Bb$ and $B$ obstructions are exactly cancelled,
leaving
$L$ to drift  round the left lobe until it meets the 
$K$ wine from the other direction, which contribution is again finite.
$$
\epsfxsize=.28\hsize\epsfbox{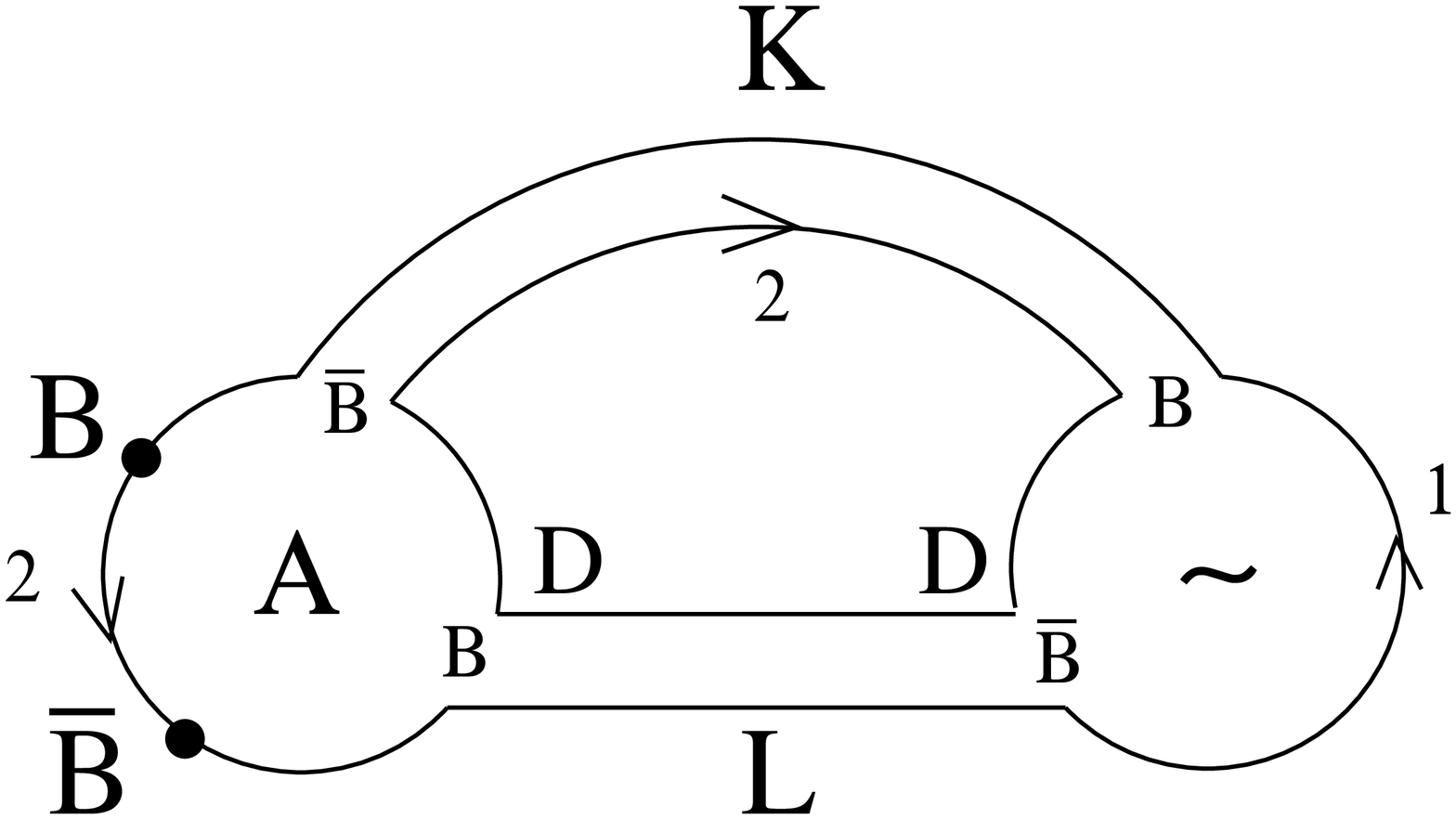}
$$
{\bf Fig.23.} A potentially divergent contribution. The $A$
in the left lobe signifies the peppered $\hS^{(1)}$ half of the
$B$ seed-action, and the $\sim$ in
the right lobe the $\ct^{-1}$ part.\hfill\break

For these reasons, it may not actually be necessary to include a
longitudinal cancelling contribution \AAL\ in the $A$--$A$ wine at all.
Here we will stick rigorously to the pattern suggested in \modprop. 
We do need $C$--$C$ contributions however, to UV 
cancel $\Bb$--$B$ $L$-type attachments to the $\ct^{-1}$ part of \kB. 
Cases where only one end attaches in this way
are cancelled with a mixed wine obtained 
by replacing one $D_\mu\cdot\delta/\delta A_\mu$ [in
the covariantization of \AAL] by $\delta/\delta C$. Then
we require a pairwise cancellation for diagrams where this mixed wine
attacks \kC.
These will follow from the $\ct^{-1}$ part of
\kB\ by replacing $\Bb_\mu$ by $D_\mu C$ and replacing one $A$ in the 
covariantization, by $\Bb$, \etc
Now consider the one-loop diagram formed as in fig.24a, where the tree-level
wine eats the gauge field in $D_\mu C$. 
$$
\epsfxsize=.8\hsize\epsfbox{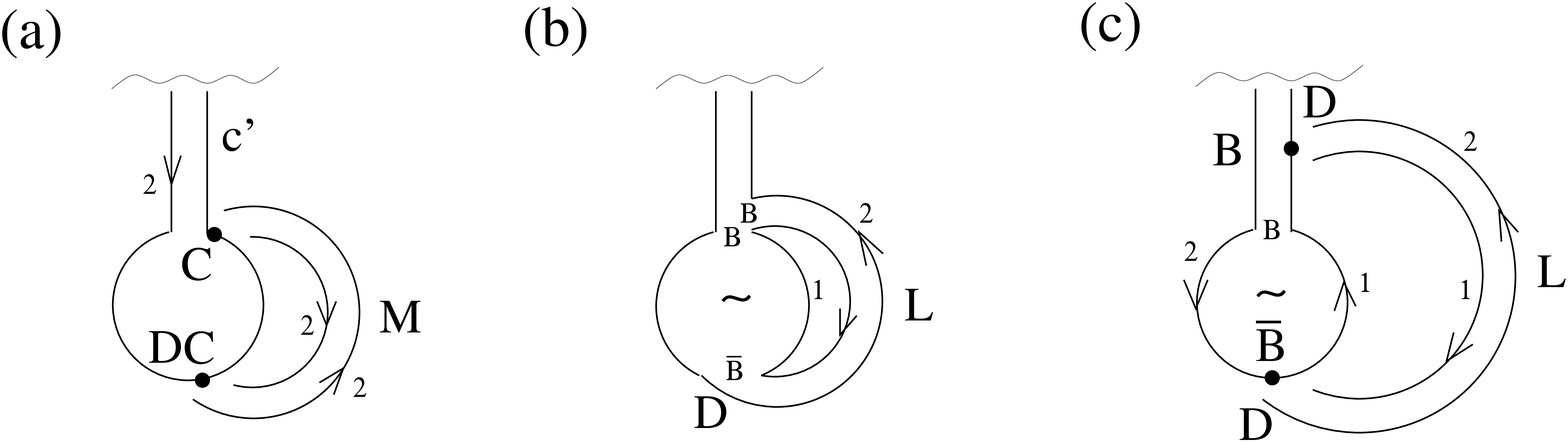}
$$
\centerline{{\bf Fig.24.} Fixing mixed wine normalisations.}
This is quadratically divergent
with a coefficient proportional to the $A$--$A$ wine 
and all that lies above it.
Its partner turns out to be fig.24b, which is 
the `back-slide' of fig.24c (using 
the effect of fig.22). These UV cancel if and
only if the second mixed $B$ wine in fig.20b is a peppered $A$--$A$ wine
(and thus in particular
normalised to $c'$). Similar arguments fix all
the other mixed wines. These choices have a wonderful side effect: the
exact cancellation of the wine pairs 
in fig.5 (where the inner loop is chosen `1', then `2').

We have worked out all the interactions necessary to implement the
idea of \modprop. UV finiteness is sufficiently obvious that it can be read
from the diagrams, and follows to all orders by pairwise
cancellations and/or by gauge invariance. It is remarkable how
little freedom is available in the choice of interactions once the
initial ideas are in place, and how at crucial points, \eg in
regularizing fig.23 or cancelling fig.5, the formalism 
saves the day. Taken together with the cancellation of
degrees of freedom, it suggests to us that we have built from
the bottom up a structure which is essentially controlled by 
supersymmetry. Given the wrong statistics of $B$, this must be 
of the Parisi-Sourlas type\ps, but given the different gauge representations
of $A^i$, $B$ and $C^i$ it cannot be normal. We suspect that if it
exists at all in the normal sense, 
it does so only on the `world-line' of the underlying
particle picture.

The classical and higher-loop solutions may now be worked out as previously.
There is an extra wrinkle however.
Consider the classical $\Bb$ $A^1$ $B$ vertex. 
We have a set of contributions to the flow in a similar way to fig.13,
which again
may immediately be integrated to give equations similar to \threep.
However this time the analogous integration constants are not determined
uniquely: instead, they may 
contain an arbitrary amount of 
\eqn\gamm{{\rm tr}\int\! \Bb_\mu F^1_{\mu\nu} B_\nu\quad,}
this vertex being
clearly consistent with the required constraints 
of $\Lambda$ independence, dimensions,
charge conjugation, Poincar\'e and gauge invariance. 
Furthermore, its undetermined
 dimensionless coefficient, call it $\gamma$, signals,
by power counting, the possibility of a log divergence in the $\Lambda$
integral $\sim\int^\infty d\Lambda_1/\Lambda_1$, proportional to \gamm.
This we indeed find, so $\gamma$ must have an infinite part in order
to cancel this and ensure a finite continuum solution.
All this in integrating over $\Lambda$, and at the classical level!
A similar situation arises with the $C$ vertices. 
The net result is that we can put $\infty$ as the upper limit, 
and take the extra parameters finite,
together with a precise prescription for throwing away
all divergences from $\Lambda$
integrations. All 
physical quantities, being universal, must anyway 
be independent of the value
of these new integration parameters. This can be verified:
for example the term proportional to $\gamma^2$ which
contributes to $\beta_1$, turns out to simplify to the total derivative 
$$\propto\int^\infty_0\!\!\!\!\!dx\, {d\over dx}
{x^2\ct^2c\,(8x\ct+c)\over(x\ct+c)^2}=0\quad.$$

There is a further subtlety in the quantum corrections, which is best
explained by a model example. Consider the $D=4$ dimensional integral
\eqn\zilch{\int\! d^4\!k\,\left\{{1\over k^2+\Lambda^2}
-{1\over|\k+\p|^2+\Lambda^2}\right\}\quad.}
These sort of cancelling divergent terms are 
generic, often as a consequence of
gauge invariance [\cf \eg the discussion of fig.23, and \gilch]. 
Expanding to $O(p^2)$, we obtain
$$\int\! d^4\!k\,\left\{ {p^2\over(k^2+\Lambda^2)^2}
-4{(\p.\k)^2\over(k^2+\Lambda^2)^3}\right\}\quad.$$
By rotation invariance we may replace 
$(\k.\p)^2$ by $k^2p^2/4$, so the above simplifies to
\eqn\pos{\Lambda^2 p^2\int\! d^4\!k\,{1\over(k^2+\Lambda^2)^3}\quad,}
a manifestly positive convergent answer, from a vanishing integral \zilch!
Where have we gone wrong? \zilch\ is finite {\sl but ambiguous}. \pos\
may be cast as a total derivative but with a finite surface term.
Generally, in situations where PV ideas are used, we must first
{\sl pre-regularise} \pv\warr. 
We do not need to subtract
with respect the pre-regulator: there are no 
overall divergences, nevertheless 
the integrals need defining carefully. Providing we preregularise in a way
that allows shifts in momenta, gauge invariance will be preserved.

Finally, we sketch the calculation of $\beta_1$. We preregularise with
dimensional regularisation: because there are no overall divergences,
a general dimension $D$ will suffice to separate
and discard total derivatives. We compute the
classical solutions up to four-point functions, directly in the continuum
(as in \threep\ and above), and substitute this into \ol\ together
with the analogous PV contributions. We keep $c(x)$ and $\ct(x)$ general 
except that $c(0)=1$ and $\ct(0)\ne0$, while
 for large momenta they go as powers, $\propto x^{-r}$
and $\propto x^{-{\tilde r}}$. Also, 
it can be shown that for full regularisation (in $D=4$) we need
$r>{\tilde r}>1$. We expand to $O(p^2)$. This gives
$2\beta_1\Delta_{\mu\nu}(p)$, defining $\beta_1$, as discussed in the 
first lecture.
We find amongst other things, many nested $\Lambda$ integrals.
Working from the inside out,
after appropriate combinations, each $\Lambda$ integral becomes a total
derivative and may be integrated exactly, and/or may be cast as a total
derivative for the $k$ integral. In this way,
the $k$ integral also
turns out to be a total derivative in $D=4$ dimensions 
(but gives garbage directly,
as above). In $D\ne4$ dimensions we thus get a
set of remainder terms proportional to $\epsilon=D-4$.
Throwing away those terms that clearly
vanish as $D\to4^-$, we found
$$\eqalign{\beta_1 &=
{N\over (4\pi)^2}{\epsilon\over2}\Bigg[
\int^\infty_0\!\!\!\!\!\!dx\ x^{\epsilon/2-1}\left\{ 2+4c^2-6c\right\}
+\int^\infty_0\!\!\!\!\!\!dx\ x^{\epsilon/2}\left\{{59\over2}{\ct\over
x\ct+c}-{59\over6}{1\over x+\sigma\ct}\right\}\cr
&\ +\int^\infty_0\!\!\!\!\!\!dx\ x^{\epsilon/2}\,
\ct \int^x_0\!\!\!\!dx_1\left\{ {20\,\ct\over (x_1\ct+c)^2}
-{2\over\ct(x_1+\sigma\ct)^2}+{24\,\ct'\over\ct(x_1\ct+c)}
-{6\,\ct'\over\ct^2(x_1+\sigma\ct)}\right\}\Bigg]\cr
&\to-{11\over3} {N\over(4\pi)^2}\ins11{as}\epsilon\to0.}$$ 

\newsec{Final Remarks}

Although we have concentrated on pure Yang-Mills, generalisation to scalars,
and chiral fermions looks straightforward. Of course $U(1)$ gauge theory
is possible and much simpler (\eg the wines are equal to their norms). It also
looks straightforward to generalise
this formalism to local coordinate invariance and
thus a \nonp\ continuum framework for
quantum gravity and supergravity.

It is important that in a regularisation that appears to generalise so
straightforwardly to in particular anomalous gauge theories, there
are subtleties in the construction. We have seen that at the quantum
level the answers are finite but well defined only after applying and
removing a preregulator.  Only the symmetries respected in this 
procedure are guaranteed preserved.
For non-anomalous theories, an imperfect pre-regulator
can be repaired by adding symmetry-breaking operators
that vanish on removing the pre-regulator,
but for anomalous theories this procedure will fail.
At the \nonp\ level we appear to have the opportunity for instability
(loss of unitarity) to percolate from the $A^2$ sector.

 We have not discussed the possible
approximation methods, apart
from sketching the underlying Wilson loop picture.
Nevertheless, we stress again that should the
present formulation allow the solution of the large $N$ limit, this
would clearly be attractive not just for the many theoretical
issues that can then be investigated, but also for the possibilities
for accurate phenomenology that such a solution may allow.

While this Wilson loop picture is clearly related to the Dyson-Schwinger
approach, there are important differences. We have seen that here
the Wilson loop is integrated over, while the gauge fields effectively
appear as background `spectators'. This in turn leads to some technical
advantages. 

At finite $N$,
Mandelstam's relations\ref\mand{S. Mandelstam, Phys. 
Rev. 175 (1968) 1580.}\LN, which
encode the overcomplete nature of the Wilson loop representation through
identities between intersecting loops, complicate the 
Dyson-Schwinger approach since these equations themselves 
are relations between intersecting loops.  Here this overcounting has no
great consequence: the fact that \eg the expansion \defWS\ of $S$
may thus integrate over the same configurations more than once (with
probably in any case zero measure) 
amounts to a harmless overparametrization of the 
effective action,\foot{Only the functional derivatives in fig.10 may 
consequently need careful definition, however they were employed only for
illustrative purposes.}
furthermore intersecting loops do not here carry 
any special significance. 

In solving the Dyson-Schwinger equations perturbatively, one finds 
a differential equation for a two-point coefficient function
$\sim < {\rm tr} A_\mu(\x) A_\nu(\y)>$, whose solution is
not unique as a consequence of gauge invariance.
In order to proceed with iterating to the higher-point functions, one
picks a solution, which amounts to fixing a gauge. At higher orders, the
corresponding ghost diagrams appear automatically \mig. 
Here there are no analogous ambiguities, no gauge fixing and no ghosts!

\bigbreak\bigskip\bigskip\bigskip\bigskip
\centerline{{\bf Acknowledgements}}\nobreak
The author would like to thank 
Tim Hollowood, Hugh Osborn and Jose Latorre for helpful conversations 
and acknowledges
support of the SERC/PPARC through an Advanced Fellowship, and PPARC grant
GR/K55738.

\listrefs

\end